\documentclass[preprint2]{aastex631}
\usepackage{longtable}

\def\dsr{$R_\odot$}

\def\dalpha{$\alpha_{\rm{MLT}}$}
\defcitealias{grev98}{GS98}

\shorttitle{Solar Models and Astrophysical $S$-factors}
\shortauthors{Yang \& Tian}

\begin{document}


\title{Solar Models and Astrophysical $S$-factors Constrained by Helioseismic Results and Updated Neutrino Fluxes}
\author[0000-0002-3956-8061]{Wuming Yang}
\affiliation{Institute for Frontiers in Astronomy and Astrophysics, Beijing Normal University, Beijing, China.}
\affiliation{School of Physics and Astronomy, Beijing Normal University, Beijing 100875, China.}
\email{yangwuming@bnu.edu.cn}

\author[0000-0003-0220-7112]{Zhijia Tian}
\affiliation{Department of Astronomy, Key Laboratory of Astroparticle Physics of Yunnan Province,
Yunnan University, Kunming 650200, China.}

\begin{abstract}
The ratio of metal abundance to hydrogen abundance of the solar photosphere, $(Z/X)_{s}$, has been revised
several times. Standard solar models, based on these revised solar abundances, are in disagreement
with seismically inferred results. Recently, Magg et al. introduced a new value for $(Z/X)_{s}$, which
is still under debate in the community. The solar abundance problem or solar modeling
problem remains a topic of ongoing debate. We constructed rotating solar models in accordance with various
abundance scales where the effects of convection overshoot and enhanced diffusion were included. Among these
models, those utilizing Magg's abundance scale exhibit superior sound speed and density profiles compared to
models using other abundance scales. Additionally, they reproduce the observed frequency separation ratios
$r_{02}$ and $r_{13}$. These models also match the seismically inferred surface helium abundance and
convection zone depth within the $1\sigma$ level. Furthermore, the calculated neutrino fluxes from these
models agree with detected ones at the level of $1\sigma$. We found that neutrino fluxes and density
profile are influenced by nuclear reactions, allowing us to use the combination of detected neutrino fluxes
and seismically inferred density for diagnosing astrophysical $S$-factors. This diagnostic approach shows
that $S_{11}$ may be underestimated by $2\%$, while $S_{33}$ may be overestimated by about $3\%$
in previous determinations. The $S$-factors favored by updated neutrino fluxes and helioseismic
results can lead to significant improvements in solar models.

\end{abstract}

\section{INTRODUCTION}
The metal abundance\texttwelveudash the mass fraction of all elements heavier than helium\texttwelveudash of a star is a key parameter
that determines the evolution and structure of the star. The estimations of the metal abundance
are related to the chemical element abundances of the Sun that are still under hot debate since
\citet{lodd03} and \citet{aspl05} revised the ratio of metal abundance to hydrogen abundance
of the solar photosphere, $(Z/X)_{s}$, from the old 0.023 \citep[hereafter GS98]{grev98} to 0.0177
or 0.0165. So far, the value of the ratio has been revised several times \citep{lodd09, aspl09,
aspl21, caff10, caff11, lodd20, amar21, magg22, desh22}. The helium abundance, $Y_{s}$, in the solar
convection zone (CZ) and thus photosphere and the radius of the base of the CZ (BCZ), $r_{\rm cz}$,
are determined by helioseismology. The surface helium abundance and metallicity inferred by \citet{voro14}
are in the range of 0.245-0.260 and 0.006-0.011, respectively. The widely accepted values of $Y_{s}$
and $r_{\rm cz}$ are $0.2485\pm0.0035$ \citep{basu04} and $0.713\pm0.003$ \dsr{} \citep{chri91} or
$0.713\pm0.001$ \dsr{} \citep{basu97}, respectively.

Standard solar models (SSMs) constructed in accordance with these revised solar abundances (low metal
abundances) disagree with the seismically inferred surface helium abundance, radius of the BCZ,
sound speed profile, and density profile \citep{bahc04b, basu04, yang07, basu09, zhang12, basu15, chri21},
as well as the detected neutrino fluxes \citep{bahc04a, turc11, sere11, yang16, yang19, yang22, zhang19}.
In order to reconcile the low-$Z$ models with helioseismology, an enhanced diffusion \citep{basu04,
mont04, guzi05} and the effects of rotation \citep{yang07, turc10, yang19, yang22}, or other effects,
such as increased opacity \citep{sere09, ayuk17, buld19, kuni21} and mass accretion of helium-poor
\citep{zhang19} or metal-poor material \citep{kuni22}, are required.

Recently, \citet{magg22} analyzed the solar photospheric abundances and obtained $(Z/X)_{s}=$ 0.0225.
This ratio is almost $8\%$ higher than the estimate by \citet{caff11}, $(Z/X)_{s}=0.0209$, mainly due
to the higher carbon, nitrogen, and neon abundances. They claimed that the puzzling mismatch between
the helioseismic constraints on the solar interior structure and the model can be resolved thanks to
this new chemical composition. Moreover, \citet{desh22} derived a photospheric solar silicon abundance
of $\log \varepsilon_{\rm Si} = 7.57\pm0.04$. Combining this with meteoritic abundances and photospheric
abundances from \citet{caff11}, they obtained $(Z/X)_{s}=0.0220\pm0.0020$, which is in agreement with
the result of \citet{magg22}. However, if the photospheric abundances of \citet{aspl21} are
used, they would obtain a smaller $(Z/X)_{s}$. Furthermore, \citet{bore22} first estimated the abundance of
C $+$ N in the Sun by using updated neutrinos. Their result is also in agreement with those of \citet{magg22}
and \citetalias{grev98}. However, the analyses of \citet{liwx21, liwx23, limc23} found that the abundances of carbon, nitrogen,
and oxygen are only marginally higher than those given by \citet{amar21}. \citet{piet23} also obtained
an oxygen abundance that is between that of \citet{amar21} and that of \citet{magg22}. \citet{buld23}
argued that ``higher metal abundances do not solve the solar problem''. Additionally, \citet{buld24} inferred
that the value of $(Z/X)_{s}$ lies in the range $0.0168-0.0205$. In fact, the surface helium abundance of
0.2439 of the \citet{magg22} model constructed in accordance with their mixtures is lower than the inferred
value of $0.2485\pm0.0035$ \citep{basu04} and that inferred by \citet{voro14}. The solar abundance problem
or solar modeling problem is still under debate.

Moreover, \citet{muss11} studied dynamic screening in solar proton–proton reactions and found that
the dynamic screening does not significantly change the reaction rate from that of the bare Coulomb potential.
However, nuclear reaction rates are a fundamental yet uncertain ingredient in constructing stellar models
\citep{bell22}. There are usually several estimated values for a nuclear cross-section factor $S(0)$
(astrophysical $S$-factor) in the literatures. For example, there exist 0.0243 \citep{bahc88}, 0.0224 \citep{bahc92},
0.0202 \citep{schr94}, 0.0274 \citep{liu96}, or 0.0208 \citep[hereafter A11]{adel11} keV barns for $S_{17}(0)$
of the $^{7}$Be($p$, $\gamma$)$^{8}$B reaction, and $3.940\times(1\pm0.004)\times 10^{-22}$ \citep{park03},
$(4.01\pm0.04)\times 10^{-22}$ \citep{adel11}, $(3.99\pm0.14)\times 10^{-22}$ \citep{chen13},
$4.047^{+0.024}_{-0.032}$ \citep{acha16}, $(4.100\pm0.037) \times 10^{-22}$ \citep{park91, acha23},
or $(4.14\pm0.07) \times 10^{-22}$ \citep{dele23} keV barns for $S_{11}(0)$ of $pp$ reaction.
The new theoretical predictions of $S_{11}(0)$ \citep{acha16, acha23, dele23} are $1-4\%$ higher than
the previously accepted value of \citet{adel11} and differ in their uncertainty estimate \citep{chen13}.
This indicates that there is an uncertainty in the theoretical calculation of $S_{11}(0)$. Moreover,
for $^{3}$He($^{3}$He, 2$p$)$^{4}$He reaction, the value of $S_{33}(0)$ is $5.0\pm0.3$ MeV barns \citep{park91},
$5.4\pm0.4$ MeV barns \citep{adel98}, $5.21\pm0.27$ MeV barns \citep{adel11} or $5.11\pm0.22$ MeV barns
at the Gamow peak. These values have a large uncertainty. The $S_{33}$ can affect $pp$I, $pp$II, and $pp$III
reactions and thus all neutrino fluxes by the feedback effect of solar luminosity calibration of solar model.
\citet{bahc04a} argued that extrapolating to the low energies to obtain $S$-factors relevant for solar
fusion introduces a large uncertainty. The factor $S_{11}$ is not accurately predicted by the chiral
effective-field-theory interactions at low chiral orders if the deuteron bound-state properties are not
adequately reproduced \citep{acha23}. \citet{bell22} showed that when other aspects of the solar model
are improved, then it shall be possible using helioseismology and solar neutrinos to improve the precision
of measurements of the nuclear cross-section factors in the $pp$ chains and CNO cycles. Helioseismology,
along with solar neutrinos, could even be utilized to determine the $S$-factors.

In this work, we mainly focus on whether the rotating and nonrotating solar models constructed in accordance with
\citet{magg22} mixtures are in agreement with seismically inferred results and updated neutrino fluxes
\citep{berg16, bore18, bore20, bore22} and what $S$-factors are favored by the seismically inferred results and
updated neutrino fluxes. The paper is organized as follows. Input physics are presented in Section 2,
calculation results are shown in Section 3, and the results are discussed and summarized in Section 4.

\section{Input Physics}

We used the Yale Rotating Stellar Evolution Code \citep{pins89, yang07, dema08} in its rotation
and nonrotation configurations to construct solar models and the \citet{guen94} pulsation code
to calculate the $p$-mode frequencies of models. We utilized the OPAL equation-of-state (EOS2005)
tables \citep{roge02} and OP opacity tables \citep{seat87, opac95, badn05, dela16}, supplemented
by the \citet{ferg05} opacity tables at low temperature, which were reconstructed with \citet{magg22}
mixtures. We employed the subroutine of \citet[hereafter B92]{bahc92} and \citet{bahc95, bahc01} to
compute the nuclear reaction rates, including neutrino fluxes. We calculated the diffusion and settling
of both helium and heavy elements by using the diffusion coefficients of \citet{thou94}.

In the atmosphere, the \citet{kris66} $T-\tau$ relation rather than the Eddington relation was adopted. Choosing
between the \citet{kris66} $T-\tau$ relation or the Eddington relation cannot change our results \citep{yang22}.
The boundary of the CZ was determined by the Schwarzschild criterion, and energy transfer by convection was treated
according to the standard mixing-length theory \citep{bohm58}. An overshoot of convection is required in order to
recover the seismically inferred depth of the CZ in our rotating models. The overshoot region was assumed to be
both adiabatically stratified \citep{chri91} and fully mixed. The depth of the overshoot region was determined
by $\delta_{\rm ov}H_{p}$ \citep{dema08}, where $\delta_{\rm ov}$ is a free parameter and $H_{p}$ is the local
pressure scale height.

In rotating models, Kawaler’s relation \citep{kawa88, chab95} was used to calculate the angular momentum
loss from the CZ due to magnetic braking, and the value of $\delta_{\rm ov}$ was 0.1. The redistribution
processes of angular momentum and chemical compositions were treated as a diffusion process \citep{enda78},
i.e.,
    \begin{equation}
       \frac{\partial \Omega}{\partial t}=f_{\Omega}
       \frac{1}{\rho r^{4}}\frac{\partial}{\partial r}(\rho r^{4}D
       \frac{\partial \Omega}{\partial r}) \,
      \label{diffu1}
    \end{equation}
for the transport of angular momentum and
    \begin{equation}
    \begin{array}{lll}
        \frac{\partial X_{i}}{\partial t}&=&f_{c}f_{\Omega}\frac{1}{\rho r^{2}}
       \frac{\partial}{\partial r}(\rho r^{2}D\frac{\partial X_{i}}
        {\partial r})\\
        & &+(\frac{\partial X_{i}}{\partial t})_{\rm nuc}-\frac{1}
       {\rho r^{2}}\frac{\partial}{\partial r}(f_{0}\rho r^{2}X_{i}V_{i}) \,
    \end{array}
      \label{diffu2}
    \end{equation}
for the change in the mass fraction $X_{i}$ of chemical species $i$, where $D$ is the diffusion
coefficient caused by rotational instabilities, including the instabilities described by \citet{pins89}
and the secular shear instability of \cite{zahn93}; $\rho$ the density; and $V_{i}$ is the velocity of
microscopic diffusion and settling given by \citet{thou94}. In the rotating models including
the effects of magnetic fields, the $D$ also includes the diffusion coefficient of magnetic fields
given by \citet{yang06}. The parameter $f_{\Omega}$ was introduced
to represent some inherent uncertainties in the diffusion equation, while the parameter $f_{c}$ was
used to account for how the instabilities mix material less efficiently than they transport angular
momentum. The default values of $f_{\Omega}$ and $f_{c}$ are $1$ and $0.03$ \citep{yang19},
respectively. The parameter $f_{0}$ is a constant. It was used to enhance the rates of diffusion
and settling, as \citet{basu04}, \citet{guzi05}, and \citet{yang07} have done, despite the fact that
there is no obvious physical justification for such a multiplier. In SSMs, the value of $f_{0}$ is
$1$, but for an enhanced diffusion model, it is larger than $1$.

All rotating and nonrotating models were calibrated to the present solar luminosity $3.844\times10^{33}$
erg s$^{-1}$, radius $6.9598\times10^{10}$ cm, mass $1.9891\times10^{33}$ g, and age $4.57$ Gyr
\citep{bahc95}. The initial hydrogen abundance $X_{0}$, metal abundance $Z_{0}$, and mixing-length
parameter \dalpha{} are free parameters adjusted to match the constraints of luminosity and radius within
around $10^{-5}$ and an observed $(Z/X)_{s}$. The initial helium abundance is determined by $Y_{0}=1-X_{0}-Z_{0}$.
The initial rotation rate, $\Omega_{i}$, of rotating models is also a free parameter adjusted to reproduce
the solar equatorial velocity of about 2.0 km s$^{-1}$. The values of these parameters are listed in
Table \ref{tab1}.

\section{CALCULATION RESULTS}

\subsection{Solar Models Constructed with the $S$-factors of B92}
\begin{figure*}
\includegraphics[angle=-90, scale=0.8]{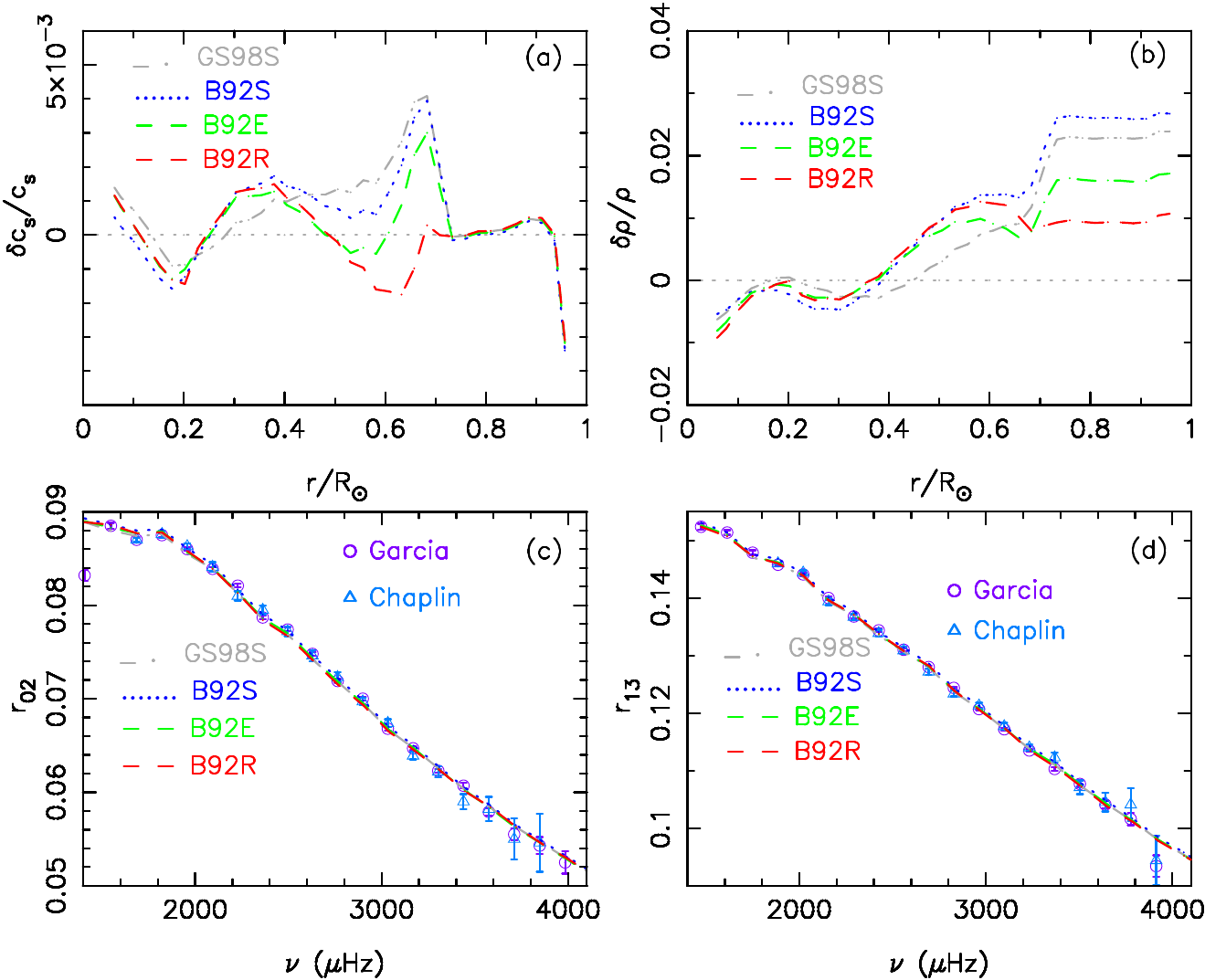}
\caption{Top panels ($a$) and ($b$): relative sound speed and density differences, in the sense (Sun-Model)/Model,
between the Sun and models. The inferred sound speed and density of the Sun are given by \citet{basu09}.
Bottom panels (c) and (d): distributions of observed and predicted ratios $r_{02}$ and $r_{13}$ as a function
of frequency. The circles and triangles show the ratios calculated from the frequencies observed
by GOLF \& VIRGO \citep{garc11} and BiSON \citep{chap99b}, respectively.
\label{fig1}}
\end{figure*}

Using the nuclear cross-section factors provided by B92, which are listed in Table \ref{tab2}, we
constructed an SSM B92S in accordance with \citet{magg22} mixtures. Tables \ref{tab1} and \ref{tab3}
show the fundamental parameters and central density and temperature of the model, respectively. Table
\ref{tab4} presents the neutrino fluxes calculated from the model. The predicted neutrino fluxes are
in agreement with those detected by \citet{bore18, bore22} except that the $^{7}$Be neutrino flux
of $4.82 \times10^{9}$ cm$^{-2}$ s$^{-1}$ is slightly lower than the detected $(4.99\pm 0.11) \times10^{9}$
cm$^{-2}$ s$^{-1}$ and the total fluxes, $8.8 \times10^{8}$ cm$^{-2}$ s$^{-1}$, of $^{13}$N, $^{15}$O,
and $^{17}$F neutrinos are slightly higher than the detected ones (see Table \ref{tab4}). The value of
$\chi_{c_{s}+\rho}^{2}$ of B92S is 811 (see Table \ref{tab5}), which is larger than 733 of the SSM GS98S
constructed in accordance with GS98 mixtures. The sound speed and density profiles of B92S are not
as good as those of GS98S (see Figure \ref{fig1}). The CZ base radius of 0.718 \dsr{} and the surface
helium abundance of 0.2400 also disagree with the seismically inferred ones. The position of the BCZ
is too shallow, and the surface helium abundance is too low. Thus, SSM B92S does not agree with helioseismic results.

The enhanced diffusion and settling can significantly improve the sound speed and density profiles of
solar models but leaves the surface helium abundance too low \citep{basu04, guzi05, yang07}.
Rotational mixing can bring the helium in the deep interior of the Sun into outer layers. The gradient
of helium abundance in the deep interior is larger than that of heavy-element abundances. The greater
the gradient, the larger the impact of rotational mixing on element abundances is. Thus, rotational
mixing can more efficiently counteract the settling of helium than of heavy-element abundances.
The low-helium problem could be resolved in rotating models. Thus we constructed an enhanced
diffusion model B92E and a rotating model B92R, in which the rates of element diffusion and settling
were increased by $15\%$ ($f_{0}=$ 1.15; see Table \ref{tab1}).

For the enhanced diffusion model B92E, the amount of the surface helium settling was increased by about $13\%$
($\approx 0.004$ by mass fraction; see Table \ref{tab1}). The surface helium abundance of $0.2373$ is too low.
However, in the rotating model B92R, rotational mixing partially counteracted the diffusion and settling
of helium and heavy elements. Thus the amount of the surface helium settling was reduced by about $33\%$
in comparison with that of B92E, which is basically consistent with the result of \citet{prof91}, who found
that macroscopic turbulent mixing can reduce the amount of the surface helium settling by around $40\%$.
As a consequence, the surface helium abundance of 0.2484 of B92R is in good agreement with the seismically
inferred value of $0.2485\pm0.0035$ \citep{basu04}. The CZ base radius of $0.711$ \dsr{} of B92R is also
consistent with the seismically inferred value of $0.713\pm0.003$ \dsr{} \citep{chri91}.

The relative sound speed difference, $\delta c_{s}/c_{s}$, and density difference, $\delta \rho/\rho$, between
the Sun and B92R are less than $0.0031$ and $0.013$, respectively. In the radiative region, the absolute value of
$\delta c_{s}/c_{s}$ is smaller than $0.0018$. The value of $289$ of $\chi_{c_{s}+\rho}^{2}$ of B92R is obviously
smaller than those of B92S and B92E (see Table \ref{tab5}). The sound speed and density profiles of B92R are
noticeably better than those of B92S, B92E, and GS98S (see Figure \ref{fig1}). Moreover, the ratios of
small to large frequency separations, $r_{02}$ and $r_{13}$ \citep{roxb03}, calculated from the theoretical
frequencies of B92R are in agreement with those computed from observed frequencies of \citet{chap99b} or
\citet{garc11} (see panels $c$ and $d$ of Figure \ref{fig1} or Table \ref{tab5}).

\begin{figure*}
\includegraphics[angle=0, scale=0.42]{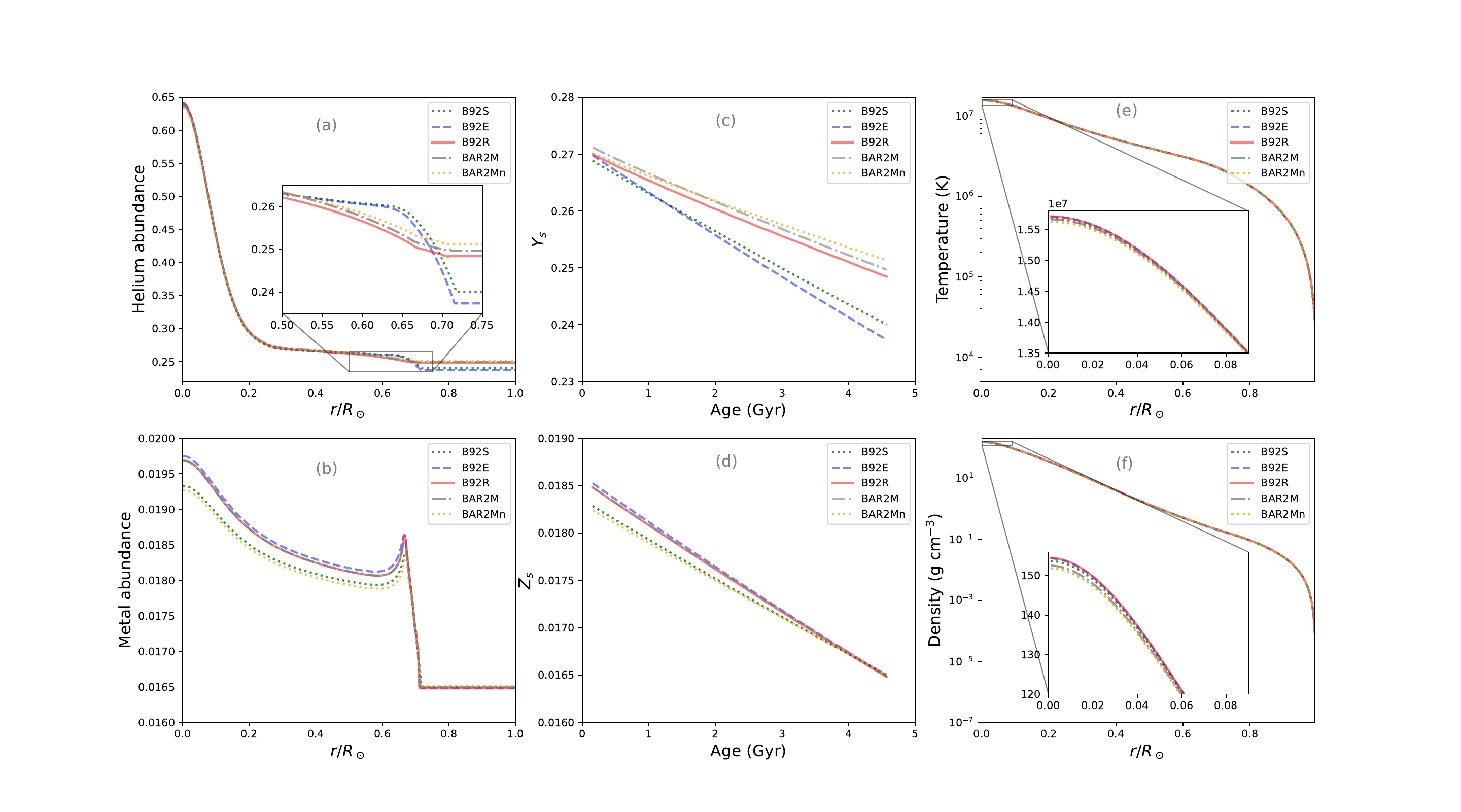}
\caption{Panels ($a$) and ($b$): distributions of helium abundance and metal abundance of different models
as a function of radius. Panels ($c$) and ($d$): surface helium and metal abundances as a
function of age. Panels ($e$) and ($f$): temperature and density as a function of radius.
\label{fig2}}
\end{figure*}

\begin{figure*}
\includegraphics[angle=0, scale=0.42]{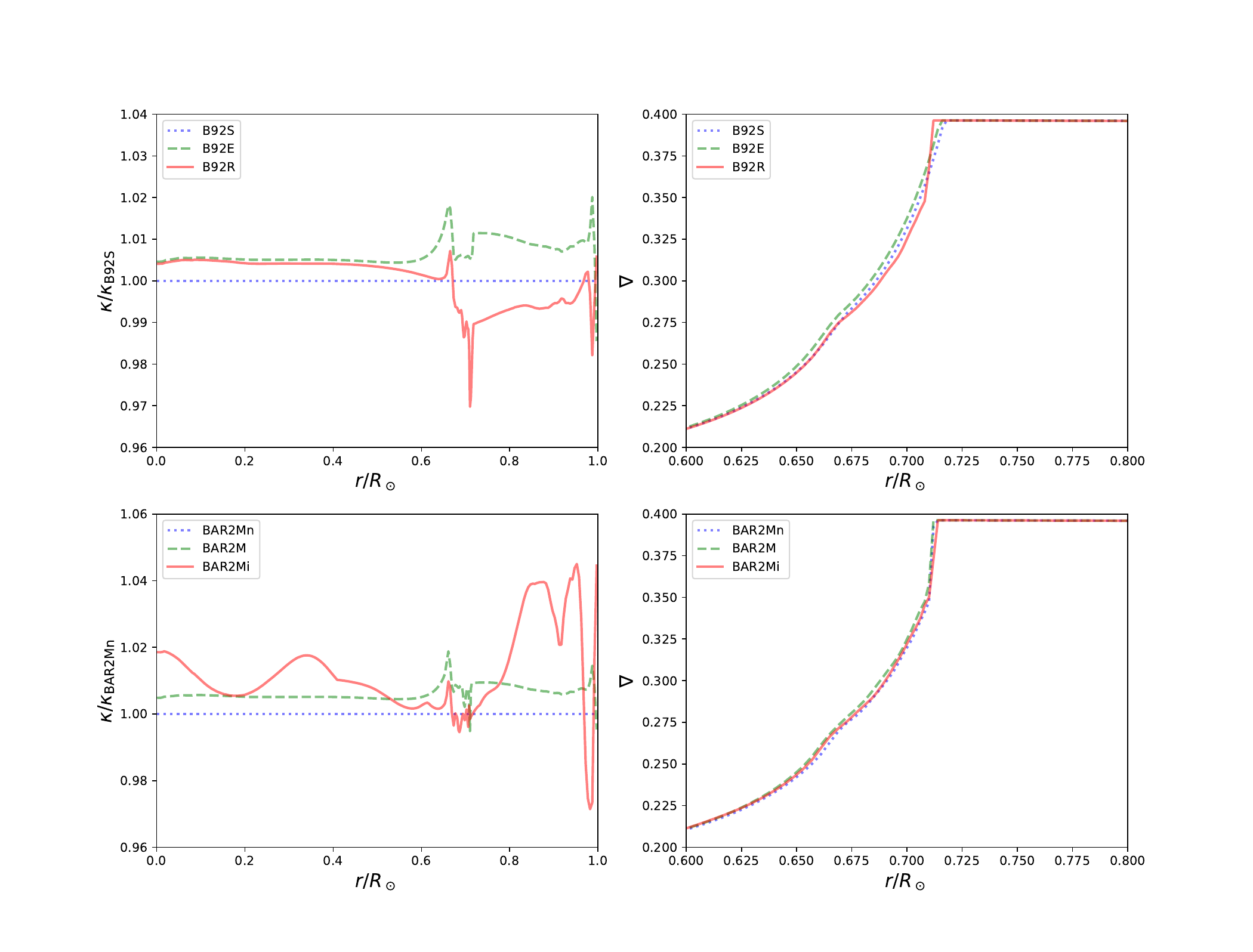}
\caption{Left panels: comparison of the Rosseland mean opacities of different models relative to that of a model.
Right panels: distributions of temperature gradient of the models as a function of radius.
\label{fig3}}
\end{figure*}

The enhanced diffusion results in B92E having a higher initial metal abundance and more metals
in the radiative region, alongside a lower helium abundance in the CZ (see Table \ref{tab1} and Figure \ref{fig2}).
The Rosseland mean opacity increases with an increase in metal abundance. For a given temperature, density,
and $Z$, it also increases with an increase in hydrogen abundance. Thus the enhanced diffusion leads to
the fact that B92E has a larger opacity and temperature gradient compared to B92S (see Figure \ref{fig3}).
The large opacity significantly improves solar models. This indicates that a higher opacity is required
in the radiative zone of low $Z$ models to reconcile the low-Z models with helioseismology, which can be
achieved by having more metals in the deep layers or by increasing the opacity itself. Rotational mixing
enhances the helium abundance and density in the CZ, while it reduces the helium abundance and thus mean
molecular weight $\mu$ around $0.6$ \dsr{} (see Figures \ref{fig1} and \ref{fig2}) compared to B92E.
The changes in the distributions of element abundances alter the opacity and temperature gradient of
rotating models. Rotating models have a steeper temperature gradient at the BCZ. The squared sound speed
is in inverse proportion to $\mu$. Thus, rotating models have a higher sound speed around $0.6$ \dsr{}.

The fluxes of $pp$, $pep$, $hep$, $^{7}$Be, and $^{8}$B neutrinos computed from B92R are in good agreement
with those detected by \citet{bore18}, while the $pp$ and $pep$ fluxes are also consistent with those
determined by \citet{berg16} at the level of $1 \sigma$ (see Table \ref{tab4}). However, the total fluxes of
$^{13}$N, $^{15}$O, and $^{17}$F neutrinos calculated from B92R are $\Phi(\rm CNO)=9.3 \times10^{8}$ cm$^{-2}$
s$^{-1}$, which are larger than the detected $6.6^{+2.0}_{-0.9} \times10^{8}$ cm$^{-2}$ s$^{-1}$ \citep{bore22}.
This could be due to overestimating the factor $S_{114}$ of $p + ^{14}$N reaction. The value of $S_{114}$
adopted by B92 is $3.32$ keV barns. But that given by \citet{angu01} and \citet{adel11} is $1.77\pm0.20$
and $1.66\pm0.12$ keV barns, respectively. When the value of $S_{114}$ is decreased from $3.32$ keV barns
to $1.77$ keV barns, the total fluxes $\Phi(\rm CNO)$ of B92R are $7.25 \times10^{8}$ cm$^{-2}$ s$^{-1}$,
which are consistent with the detected one at the level of $1\sigma$. Thus rotating model B92R is able to
reproduce the updated neutrino fluxes at the level of $1\sigma$. It is better than the SSMs GS98S
and B92S.

\begin{figure*}
\includegraphics[angle=0, scale=0.45]{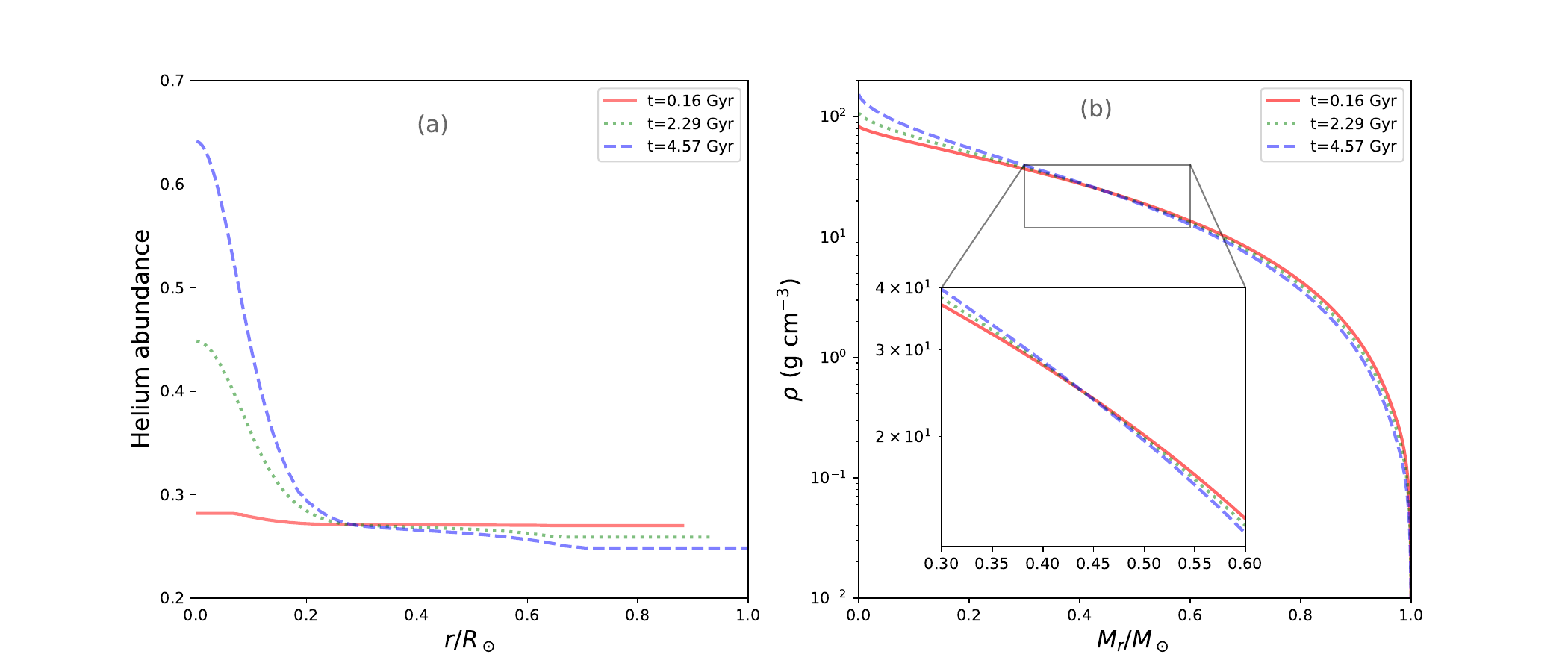}
\caption{Left panel: distribution of helium abundance of model B92R at different ages as a function of radius.
Right panel: distribution of density of model B92R at different ages as a function of mass.
\label{fig4}}
\end{figure*}

However, Figure \ref{fig1}(b) displays that the predicted density is larger than the seismically
inferred one in the central region with $r\lesssim 0.2$ \dsr{} but smaller than the inferred one
in the region with $r \gtrsim0.4$ \dsr{}. Figure \ref{fig4} shows the distributions of helium abundance and
density of model B92R at different ages as a function of radius or mass, which reveals that helium abundance
in the central region increases rapidly with an increase in age, mainly due to nuclear fusion reaction,
and that density increases with age in the central region but decreases with age in outer layers (the outer
layers expand with age). The discrepancies between the seismically inferred density and the predicted one
imply that both the contraction of core and the expansion of envelope of the models exceed what is needed to
match those of the Sun. In the central region, the change of helium abundance caused by nuclear reaction
is much larger than those deriving from element settling and rotational mixing. Additionally, the central abundances
of solar models are influenced by the feedback effect on the initial conditions of models for solar calibration.
Rotational mixing mainly affects the density profile in the CZ and at the base of the CZ (see Figures \ref{fig1}
and \ref{fig2}). The effects of element settling and mixing do not eliminate the discrepancies.

According to the definition of \citet{bahc89}, an increase in an $S$-factor must result in an increase
in fusion cross section and make the corresponding nuclear fusion reaction take place more easily
or generate more energies and neutrino fluxes under the same conditions (the same temperature and density).
Due to the fact that the luminosity of the Sun is dominated by the energy generated by the $pp$I branch
and the burning of $^{7}$Be through electron capture, an increase in $S_{11}$, $S_{33}$, and $S_{34}$
(for the $^{3}$He$+^{4}$He reaction) will lead to the fact that the solar luminosity can be reproduced by
a model with a lower temperature and density in the region with $r<0.2$ \dsr{} where nuclear reactions
take place. If the values of $S_{11}$, $S_{33}$, and $S_{34}$ are underestimated, the fusion cross sections
will be underestimated, and the solar model will need a higher density (more contraction) in the nuclear
reaction region to reproduce the solar luminosity at the age of 4.57 Gyr. The more the core contracts,
the more pronounced the expansion of outer layers becomes (see Figure \ref{fig4}(b)).
As a consequence, the density is too high in the central region but too low in outer layers.
Thus, the improper nuclear cross-section factors can lead to the density profiles of models
deviating from the seismically inferred one.

An increase in the energy released by a reaction must lead to a relative decrease in the energies
in conjunction with neutrino fluxes generated by other nuclear fusion reactions because the solar luminosity
is constant at the age of 4.57 Gyr. In $pp\rm I$ branch, the energy generated by the $^{3}$He$+^{3}$He
reaction is approximately equal to those released by $pp$ and $^{2}$H$+p$ reactions. In other words,
the luminosity $L_{pp\rm I}$ is dominated by the energy generated by the $^{3}$He$+^{3}$He reaction.
The $^{3}$He$+^{3}$He reaction does not produce any solar neutrinos. Thus increasing $S_{33}$
must lead to a decrease in the fluxes of $pp$, $pep$, $hep$, $^{7}$Be, $^{8}$B, $^{13}$N, $^{15}$O,
and $^{17}$F neutrinos and a reduction in density in the central region with $r< 0.2$ \dsr{}.
Increasing $S_{11}$ will result in an elevation in $pp$ and $pep$ neutrino fluxes and a reduction
in other neutrino fluxes and density in the central region. Similarly, an overestimate of $S_{34}$
will induce an increase in $^{7}$Be and $^{8}$B neutrino fluxes while causing a decrease in other
neutrino fluxes and density in the central region. Other $S$-factors primarily influence neutrino
fluxes directly associated with their respective $S$-factors.

The impact of overestimating $S_{33}$ on neutrino fluxes can be counteracted by the effects of
overestimating other $S$-factors. However, these overestimations inevitably result in a density
in the central region that is too low. Conversely, underestimations would lead to the density
in the central region being too high. The density profile of the Sun can be inferred through
helioseismology \citep{basu09}. These distinctive characteristics make the combination of
detected neutrino fluxes and seismically inferred density a potential tool for diagnosing
the nuclear cross-section factors. The densities in the central regions of B92S, B92E, and
B92R are lower than the inferred ones (see Figure \ref{fig1}), which may stem from underestimating
the values of $S_{11}$, $S_{33}$, and $S_{34}$.

\subsection{Solar Models Constructed with the $S$-factors of A11 }

The values of $S_{11}$, $S_{33}$, and $S_{34}$ given by \citet{adel11} are $(4.01\pm0.04)\times10^{-22}$ keV barns,
$(5.21\pm0.27)\times10^{3}$ keV barns, and $0.56\pm0.03$ keV barns, respectively, which are larger than
those adopted by B92, despite the fact that they agree with each other at the level of $1\sigma$
(see Table \ref{tab2}). Using these factors and the $S_{114}$ given by \citet{angu01}, we constructed
a rotating model A11R.

\begin{figure*}
\includegraphics[angle=-90, scale=0.8]{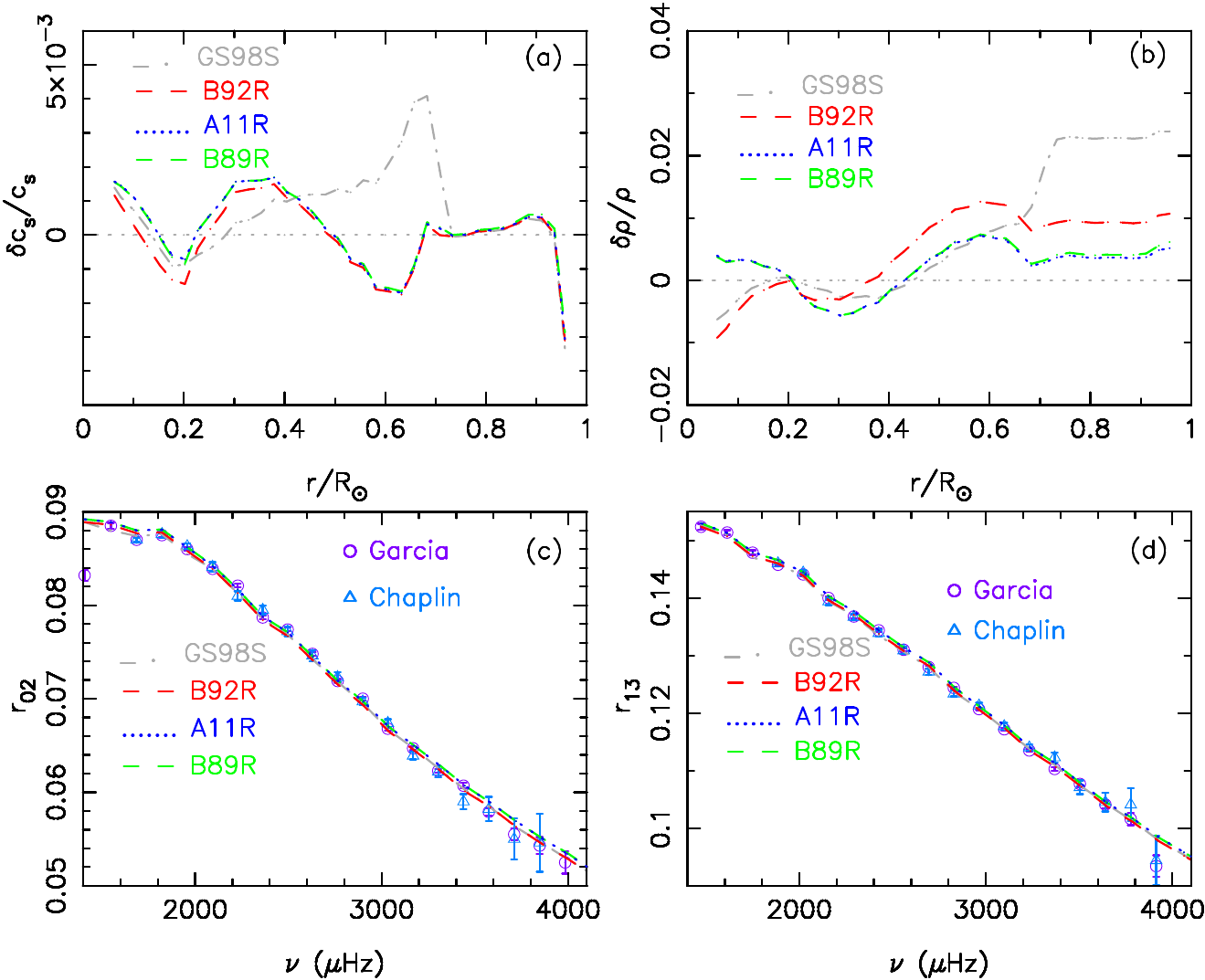}
\caption{Top panels ($a$) and ($b$): relative sound speed and density differences, in the sense (Sun-Model)/Model,
between the Sun and models. Bottom panels (c) and (d): distributions of observed and predicted ratios $r_{02}$
and $r_{13}$ as a function of frequency.
\label{fig5}}
\end{figure*}

The surface helium abundance and the CZ base radius of A11R are 0.2501 and 0.712 \dsr{}, respectively,
which agree with the seismically inferred ones at the level of $1\sigma$. Figure \ref{fig5} exhibits
the distributions of $\delta c_{s}/c_{s}$, $\delta \rho/\rho$, $r_{02}$, and $r_{13}$ of the model,
which shows that the increases in the $S$-factors ($S_{33}$ and $S_{34}$) markedly improve
the density profile, but they slightly worsen the ratios $r_{02}$ and $r_{13}$ in comparison to those
of the model B92R. Model A11R has a smaller $\chi_{c_{s}+\rho}^{2}$ but a larger $\chi_{r_{02+13}}^{2}$
than B92R (see Table \ref{tab5}). This indicates that even a small change (within
$1\sigma$) in the $S$-factors can lead to a marked change in solar models.

The total fluxes $\Phi(\rm CNO)$ predicted by A11R are $6.4 \times10^{8}$ cm$^{-2}$ s$^{-1}$,
which are in good agreement with those detected by \citet{bore22} but lower than those predicted
by B92R due to the fact that the value of $S_{114}$ given by \citet{angu01} is smaller than that
adopted by B92. However, the fluxes of $pp$ and $pep$ neutrinos predicted by A11R are obviously
lower than those determined by \citet{berg16} (see Table \ref{tab4}), despite the fact that
the fluxes are consistent with those detected by \citet{bore18} at the level of $1\sigma$.
Moreover, the flux of $5.02\times10^{6}$ cm$^{-2}$ s$^{-1}$ of $^{8}$B neutrino predicted
by the model is lower than that reported by \citet{berg16} and the one detected by \citet{bore18}.

The $pep$ reaction rate is proportional to the $pp$ reaction rate (see Equation (3.17) of \citet{bahc89}).
The fluxes of $pp$ and $pep$ neutrinos calculated from A11R are lower than those determined by
\citet{berg16}, which may result from underestimating $S_{11}$. However, if the values of $S_{33}$
and $S_{34}$ are overestimated, the $^{3}$He$ + ^{3}$He and $^{3}$He$ + ^{4}$He reactions will take place
more easily. Due to the fact that the luminosity of the Sun is constant at the age of 4.57 Gyr, the
overestimate will lead to a decrease in the energy and neutrino fluxes generated by other reactions.
Therefore, the low $pp$ and $pep$ neutrino fluxes could also derive from overestimating the values of
$S_{33}$ and $S_{34}$.

The $^{3}$He$+^{4}$He reaction produces $^{7}\rm Be +\gamma$, and then $^{7}\rm{Be}$ electron capture
produces $^{7}\rm{Li}+\nu_{e}$ and $^{7}\rm{Be}$ proton capture produces $^{8}\rm B +\gamma$. Thus,
the fluxes of both $^{7}$Be and $^{8}$B neutrinos are proportional to the ambient density of $^{7}$Be ions,
i.e. the value of $S_{34}$ can affect the fluxes of both $^{7}$Be and $^{8}$B neutrinos at the same time.
The $^{7}$Be neutrino flux of A11R is in agreement with that determined by \citet{berg16} and the one detected
by \citet{bore18}, but the $^{8}$B neutrino flux of A11R is lower than those determined by \citet{berg16}
and \citet{bore18}. These suggest that the factor $S_{34}$ could not be overestimated but the factor $S_{17}$
of $^{7}\rm{Be}$ proton capture could be underestimated by A11. As a consequence, the low $pp$ and $pep$ neutrino
fluxes could derive from underestimating the value of $S_{11}$ and overestimating the value of $S_{33}$;
and the low $^{8}$B neutrino flux could result from underestimating the value of $S_{17}$.

In order to reproduce the $pp$ and $^{8}$B neutrino fluxes of \citet{berg16}, the values of $S_{11}$
and $S_{17}$ of \citet{adel11} should be increased to about $4.13\times10^{-22}$ keV barns (an increase of about $3\%$)
and $22.4\times10^{-3}$ keV barns (an increase of about 8$\%$), respectively, which are in good agreement with
the new value of $S_{11}$ given by \citet{dele23} and the value of $S_{17}$ of \citet{bahc92}. The change
in $S_{11}$ is much larger than the uncertainty of $S_{11}$ recommended by \citet{adel11}. However, the density
in the central region with $r\lesssim0.2$ \dsr{} of this model is too low compared to the seismically inferred
one, and the $^{7}$Be and $^{8}$B neutrino fluxes calculated from this model are obviously lower than those detected
by \citet{bore18}. The helioseismic results and updated neutrino fluxes are not in favour of
$S_{11}=4.13\times10^{-22}$ keV barns and $S_{33}=5.21$ MeV barns. Reducing the value of $S_{33}$
by about $2\%$ would result in significant improvements to both the predicted neutrino fluxes and the
density profile. The value of $S_{33}$ of \citet{adel11} has a large uncertainty. Extrapolating to
the low energies to obtain $S$-factors of the solar fusion could introduce a large uncertainty.
The value of $S_{11}$ of \citet{bahc89} is larger than that of \citet{adel11} and in good agreement
with the new theoretical results of \citet{acha16, acha23} and \citet{dele23}. And the value of 5.15
MeV barns of $S_{33}$ of \citet{bahc89} is lower than that of \citet{adel11}. Thus, using the $S$-factors
of B89 (see Table \ref{tab2}), we constructed a rotating model B89R.

\subsection{Solar Models Constructed with the S-factors of B89 }

The surface helium abundance of $0.2499$ and the CZ base radius of $0.712$ \dsr{} of B89R are in good
agreement with the seismically inferred values. Figure \ref{fig5} shows that B89R exhibits superior
sound speed and density profiles (smaller $\chi_{c_{s}+\rho}^{2}$) compared to GS98S and B92R. The absolute
values of $\delta c_{s}/c_{s}$ and $\delta \rho/\rho$ between the Sun and B89R are less than $0.0029$
and $0.0074$, respectively. The values of $\chi_{c_{s}+\rho}^{2}$ and $\chi_{r_{02+13}}^{2}$ of B89R are $215$
and $2.4$, which are almost equal to those of A11R (see Table \ref{tab5}). B89R and A11R have almost
the same ratios $r_{02}$ and $r_{13}$, as well as sound speed and density profiles.

The fluxes of $pp$ and $pep$ neutrinos calculated from B89R are obviously larger than those
computed from A11R and in agreement with those determined by \citet{berg16} at the level of about $1\sigma$
(see Table \ref{tab4}). The total fluxes $\Phi(\rm CNO)$ predicted by B89R amount to $8.6 \times10^{8}$ cm$^{-2}$
s$^{-1}$, aligning with the detected $6.6^{+2.0}_{-0.9} \times10^{8}$ cm$^{-2}$ s$^{-1}$ at the $1\sigma$ level.

However, the $^{7}$Be and $^{8}$B neutrino fluxes calculated from B89R fall short of simultaneously matching
those determined by \citet{berg16} or \citet{bore18}. The $^{7}$Be neutrino flux of $4.73 \times10^{9}$ cm$^{-2}$ s$^{-1}$
of B89R is notably lower than that detected by \citet{bore18}. This can be attributed to the fact that the value
of $S_{34}$ of B89R was underestimated by about $3\%-4\%$. The $^{8}$B neutrino flux of $5.65 \times10^{6}$
cm$^{-2}$ s$^{-1}$ of B89R is in agreement with that detected by \citet{bore18} but obviously larger than the
one determined by \citet{berg16}. The predicted fluxes of both $^{7}$Be and $^{8}$B neutrinos are proportional
to the value of $S_{34}$. The increase in $S_{34}$ must lead to the fact that the predicted $^{8}$B neutrino
flux is much larger than those determined by \citet{berg16} and \citet{ahme04}. Thus, the value of $S_{17}$
of B89 could be overestimated by about $7\%$.

The value of $S_{114}$ of \citet{bahc89} is twice as much as that of \citet{adel11}. Extrapolating
to the low energies to obtain $S$-factors relevant for solar fusion could introduce a large uncertainty. Thus,
we tested both the $S$-factors of CNO cycles of \citet{bahc89} and those of \citet{adel11}. Due to the fact that
detected fluxes of $^{13}$N, $^{15}$O, and $^{17}$F neutrinos have a large uncertainty, both the CNO $S$-factors
of \citet{bahc89} and those of \citet{adel11} can reproduce the detected fluxes at the level of $1\sigma$.
The detected fluxes cannot provide a rigorous constraint on the $S$-factors of CNO cycles. But the CNO $S$-factors
of \citet{adel11} are more favored by the detected fluxes. More precise measurements are necessary to effectively
utilize neutrino fluxes for constraining the $S$-factors of CNO cycles, which could be achieved by the
Jiangmen Underground Neutrino Observatory \citep{abus23}. In the following calculations, we will adopt
the $S$-factors of CNO cycles of \citet{adel11}.

\begin{figure*}
\includegraphics[angle=0, scale=0.5]{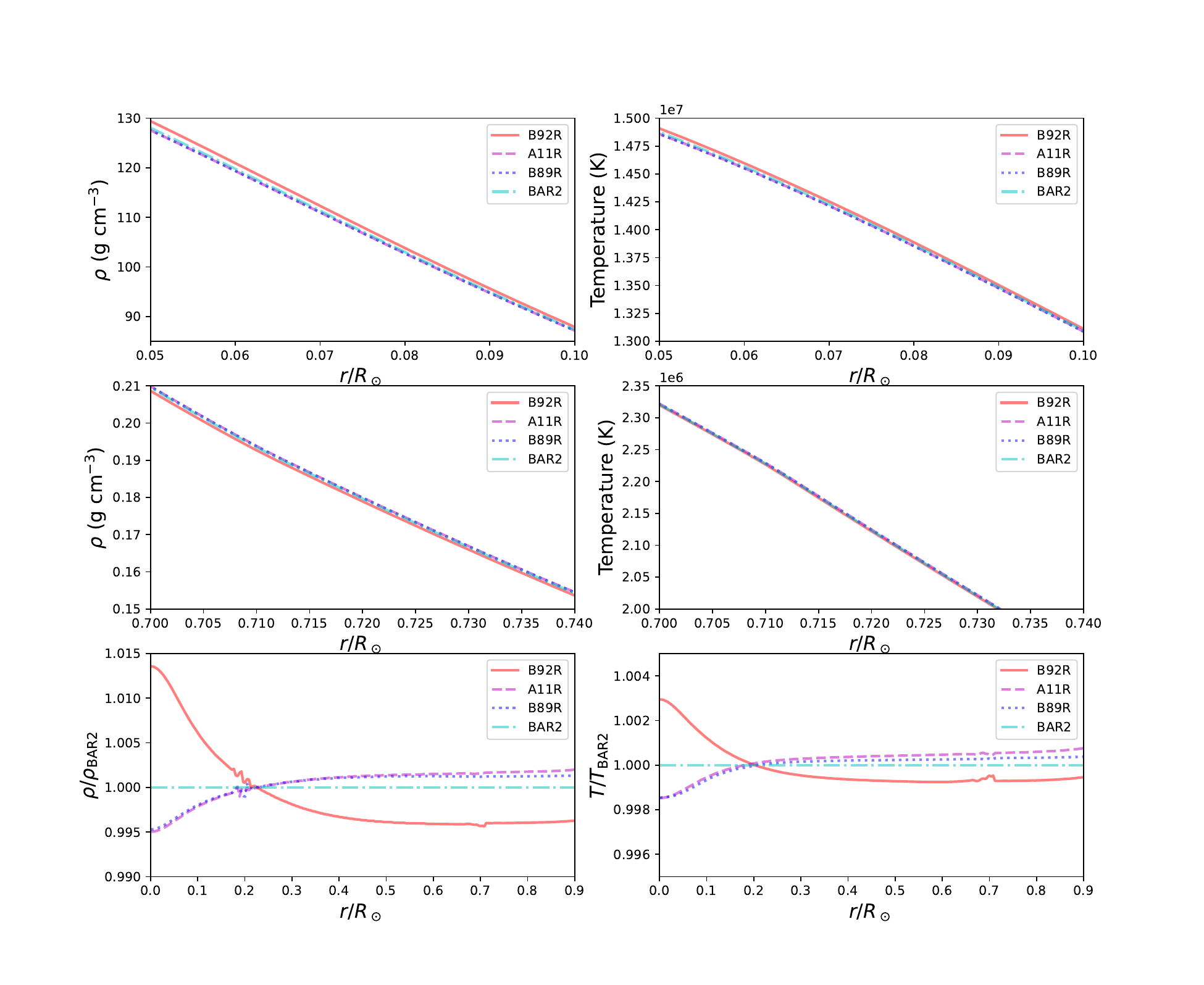}
\caption{Density and temperature distributions of solar models with different nuclear cross-section factors.
The higher the density in the central region, the lower the density in outer layers is.
\label{fig6}}
\end{figure*}

An increase in $S_{17}$ affects only the $^{8}$B neutrino flux. However, the increases in $S_{11}$, $S_{33}$,
and $S_{34}$ will lead to the fact that the solar luminosity can be reproduced by a model with a lower temperature
and density in the nuclear reaction region. Thus, the temperatures and densities of A11R and B89R are lower than
those of B92R in the central region with $r<0.2$ \dsr{} (see Table \ref{tab3} and Figure \ref{fig6}).
The lower the density in the central region, the higher the density in outer layers is,
i.e, the less the core contracts, the less pronounced the expansion of outer layers becomes
(see Figures \ref{fig4} and \ref{fig6}). Thus, the increase in $S_{11}$, $S_{33}$, and $S_{34}$ markedly
improved the density profiles of A11R and B89R in comparison to that of B92R. The density profile of B89R
is almost as good as that of A11R, but the neutrino fluxes calculated from B89R are more consistent with
the detected ones than those computed from A11R. The value of $S_{33}$ of B89 is smaller than that of A11.
This further indicates that the value of $S_{33}$ may be overestimated by A11.

\subsection{The Nuclear Cross-section Factors with the Constraints of Detected Neutrino Fluxes and
Helioseismic Results}

Model A11R shows that \citet{adel11} might overestimate the value of $S_{33}$ and underestimate
the values of $S_{11}$ and $S_{17}$. But B89R shows that \citet{bahc89} might underestimate
the value of $S_{34}$ and overestimate the value of $S_{17}$. Using $S_{33}$ of \citet{bahc89}
and $S_{17}$ of \citet{bahc92} to replace the corresponding $S$-factors of A11R (see Table \ref{tab2}),
we constructed a rotating model BAR1. In order to reproduce the $pp$ and $pep$ neutrino fluxes determined
by \citet{berg16}, the value of $S_{11}$ needs to be increased to $4.13\times10^{-22}$ keV barns, which is
consistent with the new results of \citet{dele23} and \citet{acha23} at the level of $1\sigma$.
The surface helium abundance and radius $r_{\rm cz}$ of BAR1 agree with the seismically inferred ones at
the level of $1\sigma$ (see Table \ref{tab1}). Due to the fact that the impact of the increase of $S_{11}$
on density is partially counteracted by the effect of the decrease of $S_{33}$, the changes in the $S$-factors
have a comparatively small effect on sound speed and density profiles in comparison to those of A11R.
Thus BAR1 and A11R have almost the same $\chi_{c_{s}+\rho}^{2}$ and $\chi_{r_{02+13}}^{2}$ (see Table \ref{tab5}).
But they have different neutrino fluxes (see Table \ref{tab4}) due to the fact that both a decrease
in $S_{33}$ and an increase in $S_{11}$ can lead to an increase in $pp$ and $pep$ neutrino fluxes.

\begin{figure*}
\includegraphics[angle=-90, scale=0.8]{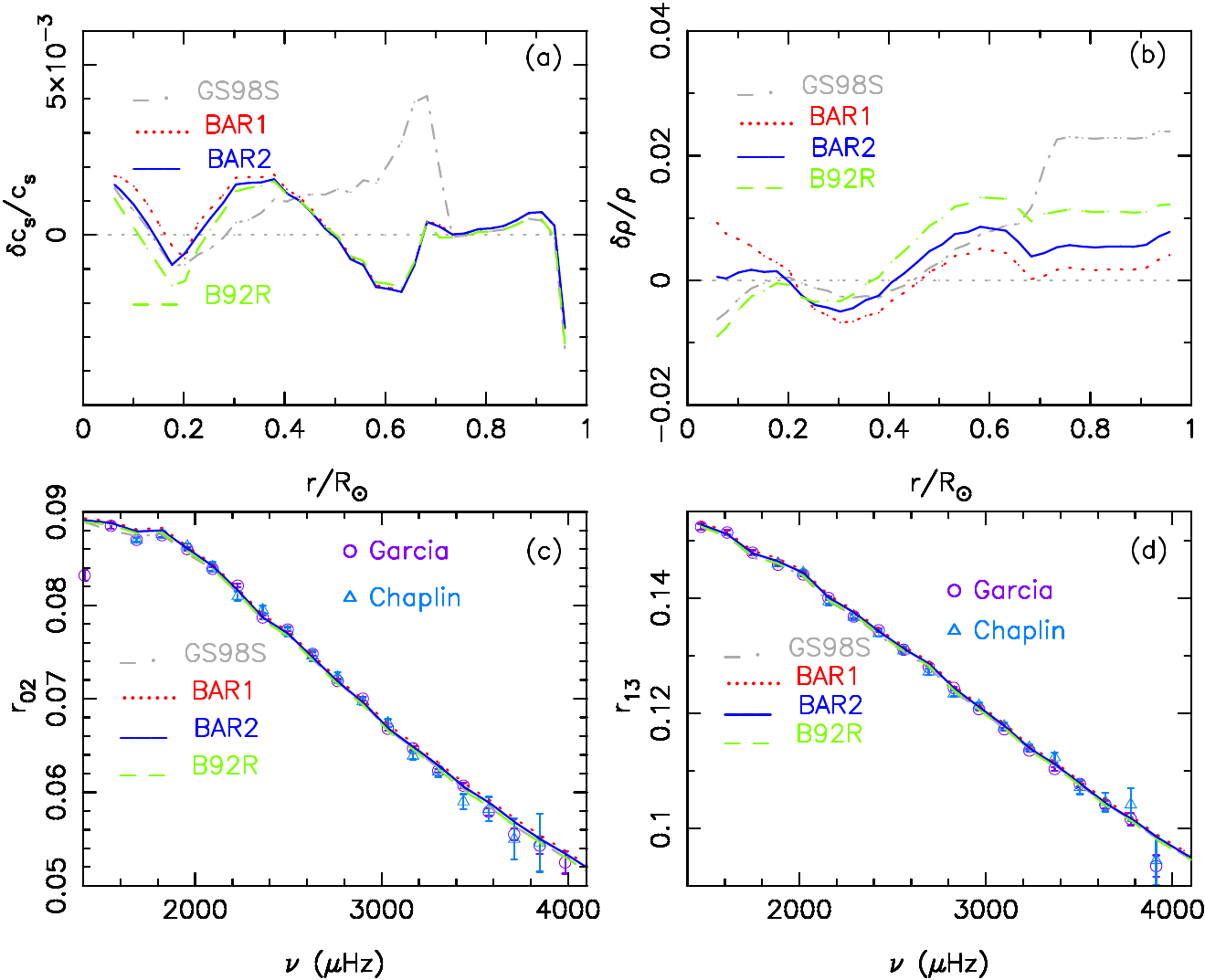}
\caption{Top panels ($a$) and ($b$): relative sound speed and density differences, in the sense (Sun-Model)/Model,
between the Sun and models. Bottom panels (c) and (d): distributions of observed and predicted ratios $r_{02}$
and $r_{13}$ as a function of frequency.
\label{fig7}}
\end{figure*}

Table \ref{tab4} shows that the fluxes of $pp$, $pep$, $^{7}$Be, and $^{8}$B neutrinos calculated from
BAR1 are consistent with those determined by \citet{berg16} at the level of $1\sigma$. The total fluxes of $^{13}$N,
$^{15}$O, and $^{17}$F neutrinos of BAR1 amount to $6.17 \times10^{8}$ cm$^{-2}$ s$^{-1}$, which are in good agreement
with the $6.6^{+2.0}_{-0.9} \times10^{8}$ cm$^{-2}$ s$^{-1}$ detected by \citet{bore18}. However,
the $^{7}$Be and $^{8}$B neutrino fluxes of BAR1 are obviously lower than those detected by \citet{bore18}.

Increasing $S_{34}$ of BAR1 alone can bring the $^{7}$Be and $^{8}$B neutrino fluxes into better
agreement with those detected by \citet{bore18} but will worsen the density profile in the central
region with $r\lesssim0.2$ \dsr{}. In addition, Figure \ref{fig7} reveals that the density in the
central region of BAR1 is lower than that inferred by \citet{basu09}. This indicates that the
values of $S_{11}$ and $S_{33}$ are not favored by helioseismic results. The values of $S_{11}$ and
$S_{33}$ are overestimated by BAR1.

Decreasing $S_{33}$ can lead to an increase in all neutrino fluxes and a rise in density
in the central region. In order to keep the $pp$ neutrino flux agreeing with the one
determined by \citet{berg16}, the value of $S_{11}$ also should be decreased at the same
time. Decreasing $S_{11}$ from $4.13\times10^{-22}$ keV barns of BAR1 to $4.07\times10^{-22}$
keV barns and decreasing $S_{33}$ from 5.15 MeV barns to 5.05 MeV barns, we constructed the rotating model BAR2.
The values of other $S$-factors of BAR2 are the same as those of BAR1 and shown in Table \ref{tab2}.
The surface helium abundance and the radius of the BCZ of BAR2 are 0.2495 and 0.712 \dsr{},
respectively, which agree with the seismically inferred ones at the level of $1\sigma$. The values
of $\chi_{c_{s}+\rho}^{2}$ and $\chi_{r_{02+13}}^{2}$ of BAR2 are 217 and 1.8, respectively.
Figure \ref{fig7} shows that BAR2 has better sound speed and density profiles than GS98S
and B92R and reproduces the distributions of ratios $r_{02}$ and $r_{13}$ calculated from observed
frequencies. The changes in $S_{11}$, $S_{33}$, and $S_{34}$ significantly improve the
density profile in the region with $r\lesssim0.2$ \dsr{} in comparison to those of B92R and BAR1.

The neutrino fluxes calculated from BAR2 are $\Phi(pp) =5.97 \times10^{10}$ cm$^{-2}$ s$^{-1}$,
$\Phi(pep) =1.441 \times10^{8}$ cm$^{-2}$ s$^{-1}$, $\Phi(hep) =9.67\times10^{3}$ cm$^{-2}$ s$^{-1}$,
$\Phi(^{7}\rm Be) =4.98 \times10^{9}$ cm$^{-2}$ s$^{-1}$, $\Phi(^{8}\rm B) =5.58 \times10^{6}$ cm$^{-2}$
s$^{-1}$, and $\Phi(\rm CNO) =6.55 \times10^{8}$ cm$^{-2}$ s$^{-1}$. These fluxes are in good agreement
with those determined by \citet{berg16} and \citet{bore18, bore22} at the level of $1\sigma$,
except for the $hep$ neutrino flux, which has a large uncertainty in detection (see Table \ref{tab4}).
Thus, the values of the $S$-factors of BAR2 are favored by helioseismic results and updated neutrino fluxes.

\subsection{Rotating Model Including the Effects of Magnetic Fields}

The surface lithium and beryllium abundances of the Sun are $1.04\pm0.10$ dex and $1.38\pm0.09$ dex \citep{lodd21},
respectively. At the time of the birth of the Sun, the Li and Be abundances are about $3.30$ and $1.48$ dex, respectively.
The solar Li depletion is about $0.6-1.2$ dex in the pre-main-sequence stage, depending on whether the convection overshoot
or another mechanism is included. There is almost no depletion for Be in the pre-main-sequence stage. The typical temperatures
for $^{7}$Li and $^{9}$Be proton-capture reaction are about $2.5\times10^{6}$ K and $3\times10^{6}$ K, respectively, which
are higher than the temperature of the BCZ of a solar model. Thus, the Li and Be elements in the CZ of an SSM cannot be
destroyed in the main-sequence stage. As a consequence, the Li abundances predicted by GS98S and B92S are about $10$ times
as large as the observed one.

\begin{figure*}
\includegraphics[angle=0, scale=0.55 ]{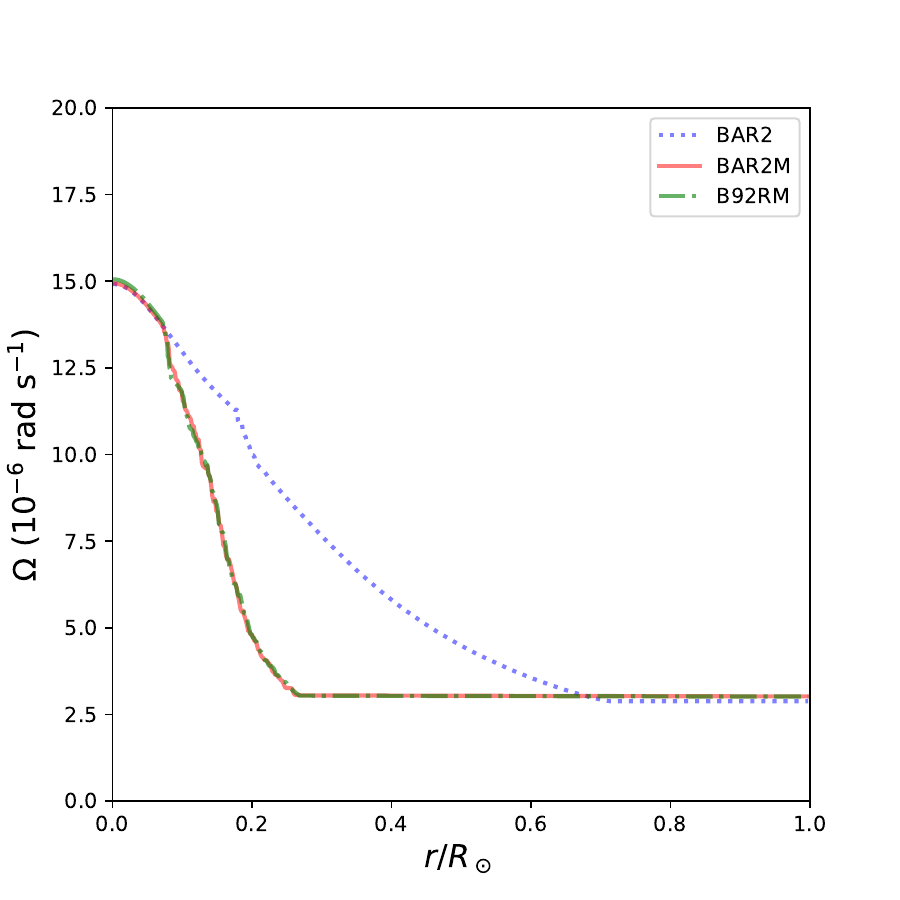}
\caption{Rotation profiles of models as a function of radius. Helioseismology shows that the Sun has a nearly flat
rotation profile above $0.2$ \dsr{} \citep{chap99a, thom03}.
\label{fig8}}
\end{figure*}

Rotational mixing can bring the helium in the radiative region into the CZ and transport the material
in the CZ into the radiative region. Thus it can simultaneously enhance the surface helium abundance and decrease
the surface Li and Be abundances. The Li and Be abundances predicted by rotating models B92R and BAR2 are 1.13
and 1.31 dex, respectively, which are consistent with the results of \citet{lodd21}. However, the rotating models
cannot reproduce the seismically inferred flat rotation profile in the external part of the radiative region (see
Figure \ref{fig8}). These indicate that B92R and BAR2 nearly mimic the material mixing processes but do not mimic
angular momentum transport of the Sun.

In order to obtain a flat rotation profile, we considered the effects of magnetic fields that were described
in detail by \citet{yang06} and \citet{yang16}, where an adjustable parameter $f_{\Omega M}$ was introduced to represent
some inherent uncertainties in diffusion equation and $f_{cM}$ was used to account for how the magnetic fields mix
material less efficiently than they transport angular momentum. With the $S$-factors of BAR2 and $f_{\Omega M}=0.001$
and $f_{cM}=2\times10^{-4}$, we constructed a rotating model BAR2M0. This model has a flat rotation profile.
But the surface helium abundance of $0.2531$ is too large, and the surface Li and Be abundances of A(Li) $=0.11$
and A(Be) $=0.98$ are too low (see Table \ref{tab1}), indicating that there is too much material exchanged between
the radiative region and the CZ of BAR2M0, i.e., the efficiency of mixing is too high.

\begin{figure*}
\includegraphics[angle=-90, scale=0.8]{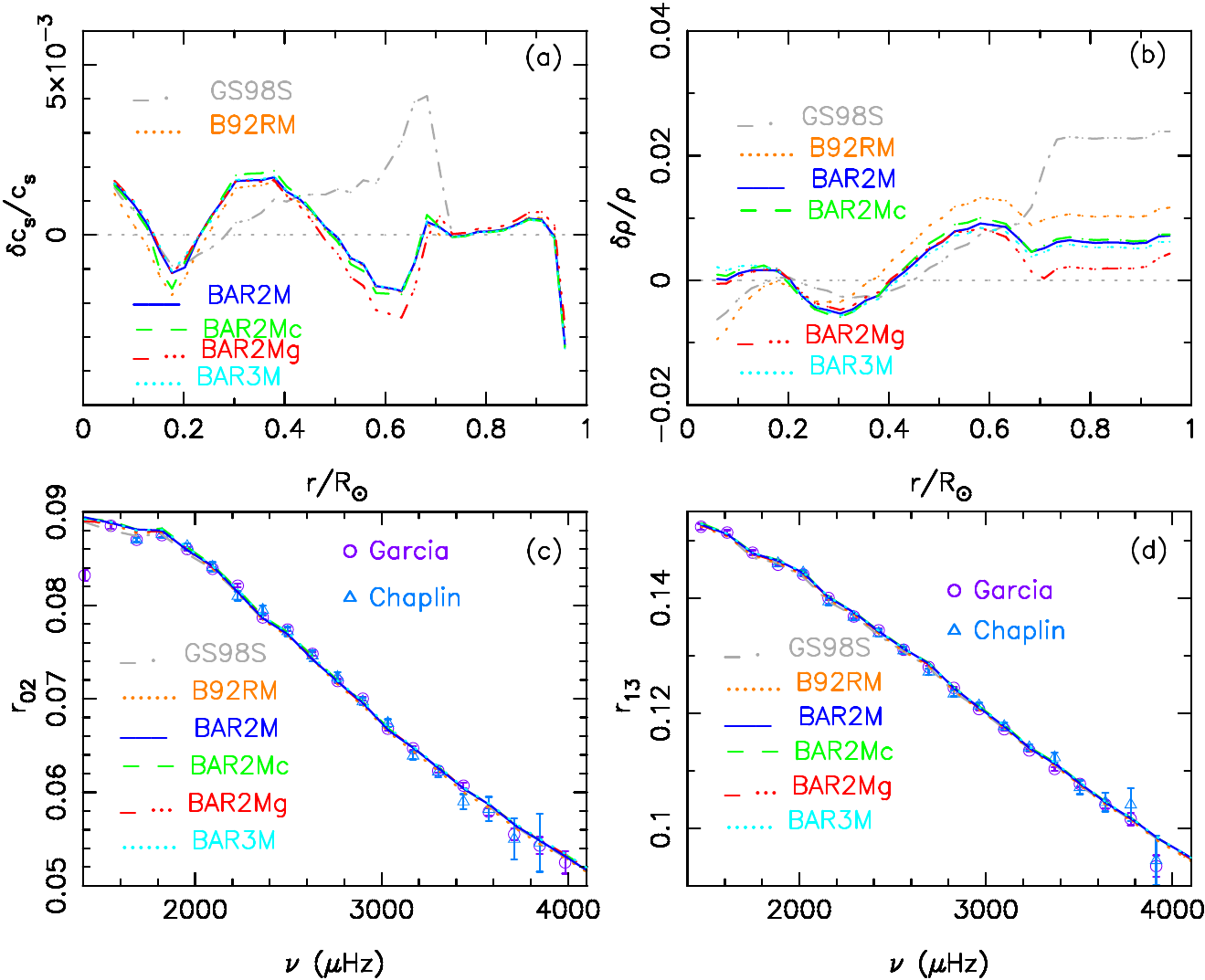}
\caption{Top panels ($a$) and ($b$): relative sound speed and density differences, in the sense (Sun-Model)/Model,
between the Sun and models. Bottom panels (c) and (d): distributions of observed and predicted ratios $r_{02}$
and $r_{13}$ as a function of frequency.
\label{fig9}}
\end{figure*}

In order to simultaneously obtain a nearly flat rotation profile and the observed Li and Be abundances,
with $f_{\Omega M}=2\times10^{-4}$ and $f_{cM}=4\times10^{-4}$, we constructed rotating models B92RM and BAR2M.
These models predict both a nearly flat rotation profile in the external part of the radiative zone and
an increase in the rotation rate in the solar core (see Figure \ref{fig8}), which are in good agreement
with the predictions of the magnetic model of \citet{egge19} and that inferred by helioseismology \citep{thom03}.
Both B92RM and BAR2M reproduce the inferred surface He, Li, and Be abundances and CZ depth (see Table \ref{tab1}).
But the sound speed and density profiles of BAR2M are in better agreement with seismically inferred ones than
those of B92RM and GS98S (see Figure \ref{fig9}). The neutrino fluxes calculated from BAR2M are almost the same
as those computed from BAR2, and are more consistent with ones determined by \citet{berg16} and \citet{bore18, bore22}
than those computed from B92RM (see Table \ref{tab4}). Thus, the effects of magnetic fields solve the problem
of the rotation profile of BAR2 but do not change the results obtained from B92R and BAR2.

The value of $4.07\times10^{-22}$ keV barns of $S_{11}$ is slightly lower than the central value of
$4.10\times10^{-22}$ keV barns given by \citet{acha23} and $4.11\times10^{-22}$ keV barns given by \citet{dele23}.
With $S_{11}=4.10\times10^{-22}$ keV barns and $S_{33}=5.05\times10^{3}$ keV barns, we constructed a rotating model BAR3M.
Other $S$-factors are the same as those of BAR2M and shown in Table \ref{tab2}. The $pp$ neutrino flux calculated from
this model is $6.00\times10^{10}$ cm$^{-2}$ s$^{-1}$. The predicted neutrino fluxes are in agreement with the detected
ones at the level of $1\sigma$ (see Table \ref{tab4}). The sound speed and density profiles of this model are
as good as those of BAR2M (see Figure \ref{fig9}). The values of $S_{11}$ supported by helioseismic data and updated
neutrino fluxes align well with recent theoretical findings \citep{acha16, acha23, dele23} but are larger than the previously
accepted value of \citet{adel11}. The increase in $S_{11}$ is much larger than the uncertainty of the $S_{11}$
recommended by \citet{adel11}. This indicates that \citet{adel11} underestimated the value of $S_{11}$ by about $2\%$
and overestimated the value of $S_{33}$ by about $2-3\%$.

An increase in $S_{33}$ would reduce the predicted $^{7}$Be neutrino flux and decrease the density in the central
region. For example, the $^{7}$Be neutrino flux calculated from a model with $S_{11}=4.10\times10^{-22}$ and
$S_{33}=5.10\times10^{3}$ keV barns is $4.89\times10^{9}$ cm$^{-2}$ s$^{-1}$. And a model having a larger $S_{33}$
generally exhibits having a larger $\chi_{r_{02+13}}^{2}$ and a lower density in nuclear reaction region.
Thus, the $^{7}$Be neutrino flux detected by \citet{bore18} and helioseismic results do not favor a model
having $S_{11}>4.10\times10^{-22}$ keV barns and $S_{33}>5.10\times10^{3}$ keV barns at the same time.

\subsection{Solar Models Constructed with Other Mixtures}

Using the $S$-factors of BAR2, we also constructed magnetic models BAR2Mc in accordance with
the \citet{caff11} mixtures and BAR2Mg in accordance with GS98 mixtures. The changes in
metallicity can affect the fluxes of $^{13}$N and $^{15}$O neutrinos (see Table \ref{tab5}).
The total fluxes of $^{13}$N, $^{15}$O, and $^{17}$F neutrinos are $\Phi(\rm CNO) =6.42 \times10^{8}$
cm$^{-2}$ s$^{-1}$ for BAR2Mc, $\Phi(\rm CNO) =6.55 \times10^{8}$ cm$^{-2}$ s$^{-1}$ for BAR2M,
and $\Phi(\rm CNO) =6.82 \times10^{8}$ cm$^{-2}$ s$^{-1}$ for BAR2Mg. The total fluxes increase with
an increase in metal abundances but they are all in agreement with the detected ones at the level of $1\sigma$.
The relative differences between the predicted fluxes are about $2\%-5\%$. If the precision of
measurements of the fluxes can reach the level of about $2\%$, the solar models constructed in
accordance with different abundance scales would be distinguished by the fluxes of $^{13}$N and $^{15}$O
neutrinos.

Figure \ref{fig9} shows that the sound speed and density profiles of BAR2M are slightly better
than those of BAR2Mc and BAR2Mg. Calculations also show that the value of $\chi_{c_{s}+\rho}^{2}$
of BAR2M is smaller than those of BAR2Mc and BAR2Mg (see Table \ref{tab5}). This indicates that
the model constructed in accordance with Magg's mixtures is slightly better than those constructed in
accordance with the \citet{caff11} or GS98 mixtures.

\section{Discussion and Summary}

In enhanced diffusion models, the velocities of diffusion and settling were increased by $15\%$. In non-rotating
models, the enhanced diffusion worsens the surface helium abundance. But in rotating models, the effect of
the enhanced diffusion on the surface helium abundance was completely counteracted by rotational mixing (see
Figure \ref{fig2}). However, we have no obvious physical justification for the increase. Moreover, the
effects of radiative accelerations on settling were not included in our models. The radiative effects
can cause the heavy element abundance and mixture to vary as a function of stellar
age and position in the Sun \citep{turco98}. The radiative acceleration effect could be counteracted by
the effects of rotation and magnetic fields.

The rotational history of the Sun is still unknown. We only considered the slow-rotating case and
the angular momentum loss mechanism of \citet{kawa88} in this work. Faster rotation and more efficient
angular momentum loss could be accompanied by more efficient material mixing that can lead to Li and Be depletion.
Different magnetic winds, such as those of \citet{matt15}, could affect the values of $f_{\Omega M}$ and $f_{cM}$
that depend on Li and Be abundances and deserve more detailed study. The rotating model of \citet{egge19}
has a larger initial velocity and a higher angular momentum loss rate than ours. The surface He abundance
of the model of \citet{buld23}, including macroscopic mixing and convection overshoot, is 0.2516, which is slightly
higher than the 0.2496 of BAR2M. The surface Li and Be abundances of the \citet{buld23} model are about 0.9
and 1.0, respectively, which are lower than the 1.1 and 1.3 of BAR2M. This indicates that the mixing of \citet{buld23}
models is more efficient than that of our models. In the deep layers, our models have a higher metal abundance
than the \citet{buld23} model. The lithium abundance of the Sun determined by \citet{wang23}
is $0.96\pm0.05$. The lithium abundances predicted by our models are $2\sigma -3\sigma$ higher than the reported
value. Additionally, the models show that there is some beryllium depletion with respect to the initial value.
This depletion could offer a more stringent constraint on material mixing during the main-sequence stage,
as beryllium depletion occurs during this phase as well.

The enhanced diffusion models have a higher initial metal abundance and more metals in the radiative region.
Thus they have a larger opacity. The rotating model BAR2Mn without enhanced diffusion has a lower initial metal
abundance and less metals in the radiative region. The values of $\chi_{c_{s}+\rho}^{2}$ and $\chi_{r_{02+13}}^{2}$
of BAR2Mn are larger than those of B92E. The calculations show that a non-rotating model with enhanced diffusion
is better than a rotating model without enhanced diffusion except the surface helium abundance. Thus a higher opacity
is required in the radiative zone of low Z models to reconcile the low-Z models with helioseismology, which can be
achieved by having more metals in the deep layers or by increasing the opacity itself. Almost the same conclusions
about the $S$-factors can also be obtained from nonrotating models with enhanced diffusion. Therefore,
the results about $S$-factors cannot be changed by the uncertainty of mixing.

\citet{kuni21} and \citet{kuni22} show that accretion with a variable composition due to planet formation
processes induces a higher central metallicity of the present-day Sun by up to $5\%$, which also can improve
solar model and predicted neutrino fluxes. The effects of the accretion is not considered in our models.
The enhanced diffusion in our models leads to an increase of about $2\%$ in the central metallicity
in comparison to that of SSM (see Table \ref{tab3}).

Figure \ref{fig6} shows the density distributions of models. The differences in the densities
have nothing to do with the variations in radii of the models. Taking a smaller radius, such as that
inferred by \citet{taka24}, mainly affects the density and sound speed above $\sim0.9$ \dsr{} and leads
to a larger $\chi_{c_{s}+\rho}^{2}$, which cannot change our results.

OP opacity is slightly larger than OPAL opacity at the base of the CZ, but by no more than $2.5\%$
\citep{badn05}. The solar models constructed by using OPAL opacity tables are not as good as those
constructed by using OP opacity tables unless OPAL opacities for the regions of the Sun with $2\times10^{6}$ K
$\lesssim T \lesssim 5\times10^{6}$ K are increased linearly by no more than $2.5\%$ centered at $T=3\times10^{6}$
K \citep{yang22}. By using this increased OPAL opacity, we constructed a magnetic model BAR2Mi
with $Z_{s}=0.0165$, $Y_{s}=0.2510$, $r_{\rm cz}=0.713$ \dsr{}, $\chi_{\nu}^{2}=0.27$, $\chi_{r_{02+13}}^{2}=1.5$,
and $\chi_{c_{s}+\rho}^{2}=206$. This model is slightly better than BAR2M. The small increase
in OPAL opacity improved solar models but did not change our results. The effects of an increase
in opacity on solar models were also studied by \citet{ayuk17} and \citet{kuni21}. They concluded that
an increase in opacity is required to reconcile the low-Z models with helioseismology. The solar model
constructed by using OPAL opacity generally has a slightly higher helium abundance in the CZ than that
constructed by using OP opacity. An increase of OP opacity in the deep layers can also lead to an increase
of helium in the CZ (see model Cop11ri of \citet{yang22}).

The fluxes of $\Phi(hep)$ predicted by models are lower than that determined by \citet{berg16}.
If the value of $S_{hep}=15.3\times10^{-20}$ keV barns \citep{wolf89} is adopted, the values
of the $hep$ fluxes predicted by BAR2M and BAR3M are about $14.3\times10^{3}$ cm$^{-2}$ s$^{-1}$,
which are in agreement with that determined by \citet{berg16} at the level of $1\sigma$. The flux of
$^{8}$B neutrino predicted by BAR2M is in good agreement with that detected by \citet{bore18} but
not consistent with that determined by \citet{berg16}. The value of $S_{17}$ given by \citet{adel11}
is $0.0208\pm 0.0016$ keV barns. If the value of $S_{17}=0.0208$ keV barns
is adopted, the $^{8}$B neutrino flux of BAR2M is $5.16 \times10^{6}$ cm$^{-2}$ s$^{-1}$,
which is in good agreement with the one determined by \citet{berg16}. The changes in $S_{hep}$
and $S_{17}$ affect only the $hep$ and $^{8}$B neutrino fluxes, respectively. They cannot affect
the sound speed and density profiles and thus cannot change our results. Thus BAR2M is able
to reproduce all neutrino fluxes at the level of $1\sigma$.

Simultaneously increasing $S_{11}$, $S_{33}$, and $S_{34}$ has almost no effect on the predicted fluxes
of $pp$, $pep$, $^{7}$Be, and $^{8}$B neutrinos. But the increases can lead to a reduction in
the predicted fluxes of $hep$, $^{13}$N, $^{15}$O, and $^{17}$F neutrinos and a decrease of density in
the central region. However, the detected fluxes of $hep$, $^{13}$N, $^{15}$O, and $^{17}$F neutrinos
cannot provide a constraint on the changes in $S_{11}$, $S_{33}$, and $S_{34}$, due to the large uncertainties
in current detections. Thus, the neutrino fluxes alone cannot be used to constrain the changes in the $S$-factors.
If the uncertainties of the fluxes of $^{13}$N, $^{15}$O, and $^{17}$F neutrinos are less than $\sim 4\%$,
they will impose a solid constraint on $S_{11}$, $S_{33}$, and $S_{34}$.

The predicted fluxes of both $^{7}$Be and $^{8}$B neutrinos are dependent on the value of $S_{34}$,
while the $^{8}$B neutrino flux also depends on $S_{17}$. They are also affected by $S_{11}$ and
$S_{33}$. An increase in either $S_{11}$ or $S_{33}$ will lead to a decrease in the fluxes of
predicted $^{7}$Be and $^{8}$B neutrinos. Models BAR1 and BAR2 share the same values for $S_{34}$
and $S_{17}$ but have different $^{7}$Be and $^{8}$B neutrino fluxes, which results from the
differences in $S_{11}$ and $S_{33}$. The fluxes calculated from BAR1 agree with those determined
by \citet{berg16}. However, if $S_{17}=20.8$ eV barns is taken, the $^{8}$B neutrino flux calculated
from BAR1 is about $4.8\times10^{6}$ cm$^{-2}$ s$^{-1}$, which is lower than that determined by \citet{berg16}.
This indicates that the $S_{11}$ and $S_{33}$ of BAR1 are too large.

The reaction rate between two nuclear species is proportional to the number densities of particles
(nuclei) and the average product of interaction cross section of the particles times velocity. The density
$\rho$ is proportional to the number densities of nuclei. The average product can be written as a formula
that is proportional to the nuclear cross-section factor $S$ (see the Equations (3.5), (3.12), and (3.13)
of \citealt{bahc89}). The solar luminosity is constant at the age of 4.57 Gyr. And the solar energy is mainly
generated by the $pp$I and $pp$II branches. The larger the nuclear cross-section factors, the more easily the
corresponding nuclear reactions occur under the same conditions. Overestimating $S_{11}$,
$S_{33}$, and $S_{34}$ would necessitate a lower density to keep the total energy generation rate
unchanged at the age of 4.57 Gyr. This would consequently lead to the density of models being lower than
the seismically inferred one in the region where nuclear fusion reactions take place. Conversely,
underestimating $S_{11}$, $S_{33}$, and $S_{34}$ would result in the density being too high.
The solar density can be constrained down to about $0.05$ \dsr{} by helioseismology. As a consequence,
the combination of detected neutrino fluxes and seismically inferred density profile in the region
with $r\lesssim0.2$ \dsr{} imposes a solid constraint on the values of $S_{11}$, $S_{33}$, and $S_{34}$.

\citet{ayuk17} concluded that the value of $S_{11}$ of \citet{adel11} should be increased by several
percent, in order to obtain a solar model with low $Z$ that is in agreement with the results of helioseismology
and detected solar neutrino fluxes. However, the updated neutrino fluxes and seismically inferred density
do not favor an $S_{11}$ larger than $4.13\times10^{-22}$ keV barns or smaller than $4.0\times10^{-22}$
keV barns and an $S_{33}$ larger than $5.15$ MeV barns or smaller than $5.00$ MeV barns.
The calculations show that the updated neutrino fluxes and seismically inferred density favor
an $S_{11}=(4.07\pm0.04)\times10^{-22}$ keV barns and an $S_{33}$ between about 5.05 and 5.10 MeV barns.
The value of the $S_{11}$ favored by helioseismic results and updated neutrino fluxes is in good
agreement with the new value of $(4.10\pm0.024\pm0.013)\times10^{-22}$ keV barns predicted by chiral effective
field theory \citep{acha23} and $4.11\times10^{-22}$ keV barns predicted by pionless effective field theory
\citep{dele23}. The increase of $S_{11}$ predicted by theories compared to previous calculations is mainly
driven by an increase in the recommended value for the axial coupling constant \citep{acha23}. The precise
$S_{11}$ from the Sun may be useful in driving down the uncertainties in the constants of the theories and
hence improve our understanding of the relevant theories \citep{bell22}.

In this work, using the $S$-factors provided by B92, A11, and B89, we constructed standard and rotating solar
models in accordance with Magg's mixtures. The surface helium abundance and the CZ depth of the SSMs do not
agree with the seismically inferred ones. The SSMs are not as good as the SSM GS98S constructed in accordance
with GS98 mixtures. In the rotating models, we included the effects of convection overshoot and enhanced diffusion.
A convection overshoot of $\delta_{\rm ov}\simeq 0.1$ is required in rotating models to recover the seismically
inferred CZ depth. The combination of enhanced diffusion and rotation brings the surface helium abundance into
agreement with the seismically inferred value at the level of $1\sigma$ and leads to rotating models having better
sound speed and density profiles than SSMs. However, in order to obtain a nearly flat rotation profile
in the external part of the radiative region (above 0.2 \dsr{}), the effects of magnetic fields are required.
As a consequence, we obtained a rotating model, B92RM, which is better than the earlier SSMs and rotating models
constructed in accordance with \citetalias{grev98} or Caffau's mixtures. Model B92RM has better sound speed and density profiles
(smaller $\chi_{c_{s}+\rho}^{2}$) than the earlier models. The surface helium abundance and radius of the BCZ
of B92RM agree with the seismically inferred ones at the level of $1\sigma$. Additionally, the neutrino fluxes
calculated from B92RM are consistent with those updated by \citet{berg16} and \citet{bore18, bore20}
at the level of $1\sigma$.

A small increase or decrease (around $1\sigma$) in nuclear cross-section factors $S_{11}$, $S_{33}$, and $S_{34}$
can result in a significant change in the predicted neutrino fluxes and density profile, particularly affecting
the density in the central region with $r\lesssim0.2$ \dsr{}. An increase in $S_{33}$ may decrease the density
in the central region and reduce the fluxes of $pp$, $pep$, $hep$, $^{7}$Be, $^{8}$B, $^{13}$N, $^{15}$O,
and $^{17}$F neutrinos. An increase in $S_{11}$ would raise the fluxes of $pp$ and $pep$ neutrinos
but decrease other neutrino fluxes and the central density. Similarly, an increase in $S_{34}$ would elevate
the fluxes of $^{7}$Be and $^{8}$B neutrinos but diminish other neutrino fluxes and the central density.
The effects of an increase in $S_{33}$ on the fluxes of $pp$, $pep$, $^{7}$Be, and $^{8}$B neutrinos can be counteracted
by the effects of an increase in $S_{11}$ and $S_{34}$. However, these increases must lead to a decrease in density in the
region where nuclear fusion reactions take place. Conversely, if $S_{11}$, $S_{33}$, and $S_{34}$ are underestimated,
the underestimation must result in the central density being too high. These distinctive characteristics make the
combination of updated neutrino fluxes and seismically inferred density profile a powerful tool for diagnosing the
$S$-factors.

Using this diagnostic approach, we found that \citet{bahc92} could underestimate the values of $S_{11}$, $S_{33}$,
and $S_{34}$ by about $2\%\textendash4\%$, while \citet{adel11} could overestimate the value of $S_{33}$ by
about $2\%\textendash3\%$ but underestimate the value of $S_{11}$ by around $2\%$. The updated neutrino fluxes
and the seismically inferred density profile favor the $S$-factors: $S_{11}=(4.07\pm0.04)\times10^{-22}$ keV barns,
$S_{33}\simeq5.05\textendash5.10$ MeV barns, $S_{34}=0.56\pm0.02$ keV barns, and $S_{17}\simeq22.4\textendash20.8$ eV barns.
These factors agree with different measurements at the level of $1\sigma$. Using these factors, we obtained a
rotating model, BAR2M or BAR3M, that is better than B92RM and the earlier models. The sound speed and
density profiles of this model are better than those of B92RM and thus better than those of the models constructed
in accordance with \citetalias{grev98} or \citet{caff11} mixtures. The surface metal abundance of BAR2M is $0.0165$. The surface
helium abundance of $0.2496$ and the radius of the BCZ of $0.712$ \dsr{} are consistent with the seismically
inferred values at the level of $1\sigma$. The initial helium abundance is $0.2722$, which is consistent with
the value of $0.273\pm0.006$ inferred by \citet{sere10}. The ratios $r_{02}$ and $r_{13}$ of BAR2M agree with
those calculated from observed frequencies. Moreover, the fluxes of $pp$, $pep$, $hep$, $^{7}$Be, and $^{8}$B
neutrinos and the total fluxes of $^{13}$N, $^{15}$O, and $^{17}$F neutrinos computed from BAR2M agree with
those determined by \citet{berg16} and \citet{bore18, bore20, bore22} at the level of $1\sigma$. This model
predicts both a nearly flat rotation profile in the external part of the radiative region and an increase
in the rotation rate in the solar core, which are in good agreement with the results of \citet{egge19}.
The results about the $S$-factors are not affected by choosing OPAL or OP opacity tables and choosing
Magg's or Caffau's mixtures.

\begin{acknowledgments}

Authors thank the anonymous referee for helpful comments that helped the authors significantly improve this work, as well
as J. W. Ferguson for providing their low-temperature opacity tables and acknowledge the support from the NSFC 11773005.
\end{acknowledgments}

\clearpage

\begin{deluxetable*}{lccccccccccccc}
\tablecaption{Fundamental Parameters of Models.
\label{tab1}}
\tablewidth{0pt}
\tablehead{
 Model &  $Y_{0}$ &  $Z_{0}$ &  \dalpha{} & $\delta_{\rm ov}$ & $f_{0}$ & $r_{\rm cz}$ & $Y_{s}$ & $Z_{s}$ & $(Z/X)_{s}$ &  $\Delta Y$
 & $\Omega_{i}$ & $A$(Li)$_{s}$ & $A$(Be)$_{s}$ }
\startdata
 GS98S$^{a}$ & 0.27562      & 0.01940 & 2.1274 & 0   & 1.0  & 0.716  & 0.2451 & 0.0174 & 0.0236 & 0.0305 & 0 & 2.14 & 1.48 \\
 B92S & 0.26994       & 0.01835 & 2.1010 & 0   & 1.0  & 0.718  & 0.2400 & 0.0165 & 0.0222 & 0.0299 & 0  & 2.15 & 1.48 \\
 B92E & 0.27107       & 0.01860 & 2.1245 & 0   & 1.15 & 0.715  & 0.2373 & 0.0165 & 0.0221 & 0.0337 & 0  & 2.15 & 1.48\\
 B92R & 0.27097       & 0.01855 & 2.0821 & 0.1 & 1.15 & 0.711  & 0.2484 & 0.0165 & 0.0224 & 0.0226 & 10 & 1.13 & 1.31\\
 \hline
 A11R & 0.27272       & 0.01857 & 2.0818 & 0.1  & 1.15 & 0.712  & 0.2501 & 0.0165 & 0.0225 & 0.0226 & 10 & 1.13 & 1.31\\
 B89R & 0.27255       & 0.01855 & 2.0818 & 0.1  & 1.15 & 0.712  & 0.2499 & 0.0165 & 0.0225 & 0.0227 & 10 & 1.13 & 1.31\\
 \hline
 BAR1 & 0.27322       & 0.01855 & 2.0833 & 0.1  & 1.15 & 0.712  & 0.2509 & 0.0165 & 0.0225 & 0.0223 & 10 & 1.13 & 1.31\\
 BAR2 & 0.27215       & 0.01855 & 2.0823 & 0.1  & 1.15 & 0.712  & 0.2495 & 0.0165 & 0.0225 & 0.0226 & 10 & 1.13 & 1.31\\
 \hline
 BAR2M0     & 0.27231 & 0.01855 & 2.0642 & 0.1 & 1.15 & 0.713 & 0.2531 & 0.0165  & 0.0226  & 0.0192 & 10 & 0.11 & 0.98\\
 B92RM      & 0.27104 & 0.01855 & 2.0804 & 0.1 & 1.15 & 0.711 & 0.2486 & 0.0165  & 0.0224  & 0.0224 & 10 & 1.09 & 1.29\\
 BAR2M      & 0.27217 & 0.01855 & 2.0806 & 0.1 & 1.15 & 0.712 & 0.2496 & 0.0165  & 0.0225  & 0.0226 & 10 & 1.08 & 1.29\\
 BAR2Mc$^{b}$ & 0.27064 & 0.01779 & 2.1016 & 0.1 & 1.36 & 0.712 & 0.2446 & 0.01548 & 0.0209  & 0.0260 & 10 & 1.05 & 1.27\\
 BAR2Mg$^{c}$ & 0.27272 & 0.01906 & 2.1147 & 0.05 & 1.15& 0.714 & 0.2501 & 0.0169  & 0.0231  & 0.0226 & 10 & 1.09 & 1.29\\
 BAR2Mn$^{d}$ &  0.27102 & 0.01830 & 2.0624 & 0.1  & 1.0 & 0.712 & 0.2513 & 0.0165  & 0.0225  & 0.0197 & 10 & 1.11 & 1.29 \\
 BAR2Mi$^{e}$ &  0.27370 & 0.01855 & 2.0700 & 0.1 & 1.15 & 0.713 & 0.2510 & 0.0165  & 0.0225  & 0.0227 & 10 & 1.10 & 1.29 \\
 BAR3M        & 0.27238  & 0.01855 & 2.0779 & 0.1 & 1.15 & 0.712 & 0.2498 & 0.0165 &  0.0225 &  0.0226 & 10 & 1.09 & 1.29 \\
  \enddata
\tablenotetext{}{Notes. The CZ radius $r_{\rm cz}$ and initial angular velocity $\Omega_{i}$ are
in units of $R_{\odot}$ and $10^{-6}$ rad s$^{-1}$, respectively. The quantity $\Delta Y=Y_{0}-Y_{s}$ is the amount
of surface helium settling. The abundance $A(\rm E)$ for an element E is defined as $A(\rm E)=12+log(n(E)/n(H))$.
The structures of these models are available at \url{
https://github.com/yangwuming/SUN/tree/main/2023sunmodels}.}
\tablenotetext{a}{The SSM is constructed by using OPAL opacity tables \citep{igle96}.}
\tablenotetext{b}{This model is constructed in accordance with \citet{caff11} mixtures, corresponding to
the model Cop11r of \citet{yang22}, but reconstructed by using different nuclear cross-section factors.}
\tablenotetext{c}{This model is constructed in accordance with GS98 mixtures.}
\tablenotetext{d}{This model is same as BAR2M but with $f_{0}=1$.}
\tablenotetext{e}{This model is same as BAR2M but constructed by using the increased OPAL opacity.}
\end{deluxetable*}

\begin{deluxetable*}{lllllll}
\tablecaption{Some of the Nuclear Cross-section Factors $S(0)$ Used in Different Models (keV barns).
\label{tab2}}
\tablewidth{0pt}
\tablehead{ Reaction                                     & B92's$^{a}$ & A11's$^{b}$ & B89's$^{c}$ & BAR1 & BAR2 & BAR3 }
\startdata
  $^{1}$H($p$, e$^{+}\nu_{e}$ )$^{2}$H ($\times10^{-22}$) & 4.00$^{+0.06}_{-0.04}$  & $4.01\pm0.04$ & 4.07 & 4.13 & 4.07 & 4.10 \\
  $^{3}$He($^{3}$He, $2p$)$^{4}$He ($\times10^{3}$)       & 5.00  & $5.21\pm0.27$   & 5.15    & 5.15    & 5.05 & 5.05  \\
  $^{3}$He($^{4}$He, $\gamma$)$^{7}$Be                    & 0.533 & $0.56\pm0.03$   & 0.54    & 0.56    &  0.56 & 0.56 \\
  $^{12}$C($p$, $\gamma$)$^{13}$N                         & 1.45  & $1.34\pm0.21 $  & 1.45    & 1.34    &  1.34 & 1.34  \\
  $^{13}$C($p$, $\gamma$)$^{14}$N                         & 5.50  &  $7.6\pm1$      & 5.50    & 7.6     &  7.6 &  7.6   \\
  $^{14}$N($p$, $\gamma$)$^{15}$O                         & 3.32  & $1.77\pm0.20^{d}$ & 3.32  & 1.77    &  1.77 &  1.77 \\
  $^{16}$O($p$, $\gamma$)$^{17}$F                         & 9.4   & $10.6\pm0.8$    & 9.4     & 10.6    &  10.6 &  10.6 \\
  $^{1}$H($p+e^{-}$, $\nu_{e}$)$^{2}$H                    & Eq.(17)$^{e}$ & Eq.(17) & Eq.(17) & Eq.(17) & Eq.(17) & Eq.(17) \\
  $^{7}$Be($e^{-}$, $\nu_{e}$)$^{7}$Li                    & Eq.(18)$^{f}$ & Eq.(18) & Eq.(18) & Eq.(18) & Eq.(18) & Eq.(18)  \\
  $^{7}$Be($p$, $\gamma$)$^{8}$B ($\times10^{-3}$)        &  22.4  & 20.8$\pm1.6$   & 24.3    & 22.4    &  22.4 &  22.4 \\
  $^{3}$He($p$, e$^{+}\nu_{e}$ )$^{4}$He ($\times10^{-20}$)& 10.35$^{g}$ & 10.35    & 10.35   & 10.35   & 10.35 & 10.35 \\
\enddata
\tablenotetext{}{Notes.}
\tablenotetext{a}{These factors are given in \citet{bahc92}.}
\tablenotetext{b}{Given by \citet{adel11}.}
\tablenotetext{c}{Given by \citet{bahc88} and \citet{bahc89}. }
\tablenotetext{d}{This value is given by \citet{angu01}; the value of $S_{114}(0)$ given by \citet{adel11} is $1.66\pm0.12$.}
\tablenotetext{e}{The Eq.(3.17) of \citet{bahc89}, but the factor of 1.102 was replaced by the factor of 1.130 \citep{adel11}.}
\tablenotetext{f}{The Eq.(3.18) of \citet{bahc89}, but the factor of 5.54 was replaced by the factor of 5.60 \citep{adel11}.}
\tablenotetext{g}{The value of $S_{13}(0)$ given by \citet{schi92} is $2.30\times10^{-20}$, which should be multiplied
by a factor of $4.5$ \citep{marc20}; the value is $(8.6\pm2.6)\times10^{-20}$ in \citet{adel11} or $8\times10^{-20}$ in \citet{bahc89}.}
\end{deluxetable*}

\begin{deluxetable*} {lcccc}
\tablecaption{The Central Temperature, Density, Helium Abundance, and Metallicity of Models.
\label{tab3}}
\tablehead{
\colhead{Model} & \colhead{$T_{c}$} & \colhead{$\rho_{c}$} & \colhead{$Y_{c}$} & \colhead{$Z_{c}$}
}
\startdata
 GS98S& 15.770 & 154.2 & 0.6450 & 0.02045 \\
 B92S & 15.676 & 153.7 & 0.6381 & 0.01933 \\
 B92E & 15.714 & 154.4 & 0.6416 & 0.01975 \\
 B92R & 15.712 & 154.5 & 0.6412 & 0.01969 \\
 \hline
 A11R & 15.643 & 151.7 & 0.6360 & 0.01971 \\
 B89R & 15.643 & 151.7 & 0.6360 & 0.01969 \\
 \hline
 BAR1 & 15.615 & 150.6 & 0.6340 & 0.01969 \\
 BAR2 & 15.666 & 152.5 & 0.6378 & 0.01969 \\
 \hline
 BAR2M0 & 15.668 & 152.7 & 0.6383 & 0.01969 \\
 B92RM  & 15.713 & 154.6 & 0.6415 & 0.01969 \\
 BAR2M  & 15.666 & 152.6 & 0.6380 & 0.01969 \\
 BAR2Mc & 15.687 & 152.4 & 0.6396 & 0.01909 \\
 BAR2Mg & 15.680 & 152.8 & 0.6387 & 0.02023 \\
 BAR2Mn & 15.629 & 151.9 & 0.6345 & 0.01927 \\
 BAR2Mi & 15.693 & 152.5 & 0.6395 & 0.01969 \\
 BAR3M  & 15.654 & 152.1 & 0.6370 & 0.01969 \\
\enddata
\tablenotetext{}{Notes. The central temperature $T_{c}$ and density $\rho_{c}$ are in
units of $10^{6}$ K and g cm$^{-3}$, respectively.}
\end{deluxetable*}

\begin{deluxetable*} {lcccccccc}
\tablecaption{Measured and Predicted Solar Neutrino Fluxes ($\mathrm{cm}^{-2}\ \mathrm{s}^{-1}$).
\label{tab4}}
\tablehead{
\colhead{Model} & \colhead{$pp$} & \colhead{$pep$} &\colhead{$hep$} &\colhead{$^{7}$Be} & \colhead{$^{8}$B} & \colhead{$^{13}$N} & \colhead{$^{15}$O} &\colhead{$^{17}$F}  \\
\colhead{} & \colhead{$\times10^{10}$} & \colhead{$\times10^{8}$} &\colhead{$\times10^{3}$} &\colhead{$\times10^{9}$} & \colhead{$\times10^{6}$} & \colhead{$\times10^{8}$} & \colhead{$\times10^{8}$} &\colhead{$\times10^{6}$}
}
\startdata
 Measured &  6.06$^{+0.02 a}_{-0.06}$ & 1.6$\pm$0.3$^{b}$ &...& 4.84$\pm$0.24$^{a}$& 5.21$\pm$0.27$^{c}$ &...& ...&...\\
 B16$^{d}$ &  5.97$^{+0.04}_{-0.03}$ & 1.448$\pm$0.013 & 19$^{+12}_{-9}$ & 4.80$^{+0.24}_{-0.22}$ & 5.16$^{+0.13}_{-0.09}$ & $\leq$13.7 & $\leq$2.8 & $\leq$85 \\
Borexino$^{e}$ & 6.1$\pm$0.5 & 1.39$\pm$0.19 & $<$220 & 4.99$\pm$0.11 & 5.68$^{+0.39}_{-0.41}$ & \multicolumn{2}{c}{6.6$^{+2.0}_{-0.9}$} &  \\
\hline
\hline
 GS98S  &  5.92 & 1.428 & 9.65 & 5.02 & 6.01 & 5.64 & 4.91 & 4.51  \\
 B92S   &  5.96 & 1.449 & 9.82 & 4.82 & 5.45 & 4.70 & 4.06 & 5.13  \\
 B92E   &  5.95 & 1.444 & 9.76 & 4.91 & 5.68 & 4.95 & 4.31 & 5.45  \\
 B92R   &  5.96 & 1.448 & 9.76 & 4.90 & 5.66 & 4.94 & 4.29 & 5.43  \\
 \hline
 A11R   &  5.86 & 1.409 & 9.64 & 4.91 & 5.02 & 4.24 & 2.10 & 5.60  \\
 B89R   &  5.95 & 1.432 & 9.64 & 4.73 & 5.65 & 4.58 & 3.93 & 4.96  \\
 \hline
 BAR1   &  5.98 & 1.435 & 9.59 & 4.81 & 5.20 & 4.10 & 2.02 & 5.39  \\
 BAR2   &  5.97 & 1.441 & 9.67 & 4.98 & 5.58 & 4.33 & 2.16 & 5.75  \\
 \hline
 BAR2M0 &  5.98 & 1.444 & 9.67 & 4.98 & 5.58 & 4.34 & 2.16 & 5.77  \\
 B92RM  &  5.96 & 1.448 & 9.76 & 4.90 & 5.66 & 4.94 & 4.30 & 5.43  \\
 BAR2M  &  5.98 & 1.442 & 9.67 & 4.98 & 5.58 & 4.33 & 2.16 & 5.76  \\
 BAR2Mc &  5.97 & 1.441 & 9.68 & 5.01 & 5.69 & 4.24 & 2.12 & 5.67  \\
 BAR2Mg &  5.97 & 1.441 & 9.65 & 5.01 & 5.66 & 4.51 & 2.25 & 6.02  \\
 BAR2Mn &  5.99 & 1.446 & 9.73 & 4.89 & 5.35 & 4.10 & 2.03 & 5.42 \\
 BAR2Mi &  5.97 & 1.439 & 9.65 & 5.03 & 5.71 & 4.41 & 2.20 & 5.89 \\
 BAR3M  &  6.00 & 1.447 & 9.65 & 4.94 & 5.48 & 4.28 & 2.13 & 5.67 \\
\enddata
\tablenotetext{a}{\citet{bell11}.}
\tablenotetext{b}{\citet{bell12}.}
\tablenotetext{c}{\citet{ahme04}.}
\tablenotetext{d}{\citet{berg16}.}
\tablenotetext{e}{\citet{bore18}. The total fluxes, $\Phi(\rm CNO)$, produced by CNO cycle are given by \citet{bore22}.}
\end{deluxetable*}

\begin{deluxetable*} {lcccc}
\tablecaption{The Values of $\chi^{2}$ of Different Models. \label{tab5}}
\tablehead{
 Model   &  $\chi_{c_{s}+\rho}^{2\quad \ a}$  & $\chi^{2\quad \ \ b}_{r_{02+13}}$   & $\chi^{2\qquad \ c}_{\rm neutrino}$
 & $\chi^{2\quad \ \ d}_{\rm helium}$
}
\startdata
GS98S  & 733  & 1.2  & 1.83  & 0.94 \\
B92S   & 811  & 2.3  & 0.85  & 5.90 \\
B92E   & 376  & 1.4  & 0.66  & 10.24\\
B92R   & 289  & 1.5  & 0.61  & 0.00\\
\hline
A11R   & 219  & 2.4  & 4.45  & 0.21 \\
B89R   & 215  & 2.4  & 1.60  & 0.16 \\
\hline
BAR1   & 220  & 3.3  & 1.07  & 0.36 \\
BAR2   & 217  & 1.8  & 0.24  & 0.08 \\
\hline
 BAR2M0 & 251 & 2.0 & 0.22 & 1.73 \\
 B92RM  & 310 & 1.4 & 0.61 & 0.0 \\
 BAR2M  & 237 & 2.0 & 0.24 & 0.10 \\
 BAR2Mc & 285 & 2.3 & 0.24 & 1.24 \\
 BAR2Mg & 289 & 1.7 & 0.24 & 0.24 \\
 BAR2Mn & 404 & 3.7 & 0.50 & 0.64 \\
 BAR2Mi & 206 & 1.5 & 0.28 & 0.51 \\
 BAR3M  & 229 & 2.1 & 0.35 & 0.14 \\
\enddata
\tablenotetext{}{Notes. The function $\chi^{2}$ is defined as $\chi^{2}=\frac{1}{N}\sum^{N}_{i=1}
\frac{(q_{\rm ob, i}-q_{\rm th, i})^{2}}{\sigma^{2}_{i}}$, where $q_{\rm ob,i}$ and $q_{\rm th,i}$ are
the observed/inferred and theoretical values of quantities $q_{\rm i}$, respectively;
$\sigma_{i}$ are the errors associated to the corresponding observed/inferred quantities; $N$ is the number of the quantities.}
\tablenotetext{a}{The inferred $c_{s}$ and $\rho$ are given in \citet{basu09}.}
\tablenotetext{b}{The observed $r_{02}$ and $r_{13}$ are calculated from the frequencies given by \citet{chap99b}.}
\tablenotetext{c}{The values of $\chi^{2}_{\rm neutrino}$ were calculated by using the fluxes $\Phi(pp)$, $\Phi(pep)$,
and $\Phi(hep)$ determined by \citet{berg16} and $\Phi(\rm Be)$, $\Phi(\rm B)$, and $\Phi(\rm CNO)$ detected by
\citet{bore18, bore22}. }
\tablenotetext{d}{The inferred helium is $0.2485\pm0.0035$ \citep{basu04}.}
\end{deluxetable*}

\end{document}